\documentclass[useAMS,usenatbib]{mn2e}

\newcommand{\kms}{ km s$^{-1}$\xspace}
\newcommand{\ang}{\mbox{\AA}\xspace}
\newcommand{\nhI}{N$_{\rm H \ I}$\xspace}
\newcommand{\w}{W$_{0}$}
\newcommand{\za}{z$_{abs}$}
\newcommand{\lala}{$\lambda\lambda$\xspace}

\usepackage{graphicx}
\usepackage{times}
\usepackage{colordvi}
\usepackage{epsfig}
\usepackage{lscape}
\usepackage{rotating}
\usepackage{xspace}
\begin{document}

\title[Sub-DLAs and strong Lyman-limit systems at z $\la$ 1.5.]{The chemical compositions of 10 new sub-DLAs and strong Lyman-limit systems at z $\la$ 1.5. }
\author[J. D. Meiring, V.P. Kulkarni et al.]{Joseph D. Meiring$^{1}$, 
Varsha P. Kulkarni$^{1}$, 
James T. Lauroesch$^{2}$,
Celine P\'eroux$^{3}$,
\newauthor  Pushpa Khare$^{4}$, Donald G. York$^{5,6}$, $\&$ Arlin P. S. Crotts$^{7}$\\ 
$^{1}$Department of Physics and Astronomy, University of South Carolina, Columbia, SC 29208, USA \\
$^{2}$Department of Physics and Astronomy, University of Louisville, Louisville, Ky 40292 USA\\
$^{3}$Observatoire Astronomique de Marseille-Provence, Traverse du Siphon, Marseille, France \\
$^{4}$Department of Physics, Utkal University, Bhubaneswar, 751004, India \\
$^{5}$Department of Astronomy and Astrophysics, University of Chicago, Chicago, IL 60637, USA \\ 
$^{6}$Enrico Fermi Institute, University of Chicago, Chicago, IL 60637, USA \\
$^{7}$Department of Astronomy, Columbia University, New York, NY 10027, USA\\}

\date{Accepted ... Received ...; in original form ...}

\pagerange{\pageref{firstpage}--\pageref{lastpage}} \pubyear{}

\maketitle

\label{firstpage}

\begin{abstract}
We present chemical abundance measurements from medium resolution observations of 8 sub-damped Lyman-$\alpha$ absorber and 2 strong Lyman-limit systems at z $\la$ 1.5 observed with 
the MIKE spectrograph on the 6.5m Magellan II Clay telescope. These observations were taken as part of an ongoing project
 to determine abundances in \za $\la$ 1.5 quasar absorption line systems (QSOALS) focusing on sub-DLA systems. 
These observations increase the sample of Zn measurements in \za $\la$ 1.5 sub-DLAs by $\sim$50$\%$.
Lines of Mg I, Mg II, Al II, Al III, Ca II, Mn II, Fe II, and Zn II were detected and column densities were determined. Zn II, a relatively undepleted element and tracer of the gas phase 
metallicity is detected in two of these systems, with [Zn/H]$=-$0.05$\pm$0.12 and [Zn/H]$>$+0.86. 
The latter system is however a weak system with \nhI$<$18.8, and therefore may need significant ionisation corrections to the abundances.  
Fe II lines were detected in all systems, 
with an average Fe abundance of $<$[Fe/H]$>$$=-$0.68, higher than typical Fe abundances for DLA systems at these redshifts. 
This high mean [Fe/H] could be due to less depletion of Fe onto dust grains, 
or to higher abundances in these systems. We also discuss the relative abundances in these absorbers. The systems with high metallicity  show high ratios of [Mn/Fe] and [Zn/Fe],  
as seen previously in another sub-DLA. These higher values of [Mn/Fe] could be a result of heavy depletion of Fe onto grains, unmixed gas, or an intrinsically non-solar abundance
pattern. Based on Cloudy modeling, we do not expect ionisation effects to cause this phenomenon. 

\end{abstract}


\section{Introduction}
There are still many open questions concerning the processes of galaxy formation and evolution. Measurements of the abundances of heavy elements
 in galaxies give important information
about the ongoing processes of star formation and death and the overall chemical enrichment of these galaxies. Studying galaxies at higher redshifts through emission is 
often difficult and time consuming, and biases towards more luminous galaxies are often present. Quasar absorption line systems (QSOALS) provide a means of studying
the interstellar medium (ISM) of high redshift galaxies independent of the  morphology of galaxies (that is, not based on the selection of galaxies of certain 
morphology). In addition, absorption lines in QSO spectra allow us to study the diffuse intergalactic medium using lines of O VI (e.g., \citealt{Sim02}), or using X-Ray absorption 
lines of more highly ionized species \citep{Fang02}. 

 Quasar absorption line systems with strong Lyman-$\alpha$ lines are often divided into two classes: Damped Lyman-$\alpha$ (DLAs, log N$_{\rm H \ I}$ $\ge$ 20.3) and 
 sub-Damped Lyman-$\alpha$ (sub-DLA 19 $\la$ log N$_{\rm H \ I}$ $<$ 20.3, \citealt{Per01}) are expected to contain a major fraction of the neutral gas in the Universe, 
 while the majority of the baryons are thought to lie in the highly ionized and diffuse Lyman-$\alpha$ forest clouds with log N$_{\rm H \ I}$ $\la$ 14 in intergalactic space
 \citep{Petit93}. The DLA and sub-DLA systems are generally believed to be associated directly with galaxies, at all redshifts in which they are seen. 
 Observations of the galaxies in emission responsible for the absorption line systems have been met with mixed results. In several instances, a galaxy at the same redshift in emission 
 could not be found near the line of sight of the QSO (see for example \citealt{Kul00, Kul01, Chen03, Rao03}.) 

 DLAs have been the preferred systems for chemical abundance studies thanks to their high gas content. 
 However, most DLAs are found to be metal poor, typically far below the solar level (e.g., \citet{Kul05} and references therein).  
 Based on Fe II abundance measurements, \citet{Per03a} suggested stronger metallicity evolution in sub-DLA systems than DLA systems. 
 Specifically, \citet{Per03a} showed that the \nhI-weighted mean Fe metallicity in sub-DLAs increased from $\sim$ 1/100 Z$_{\sun}$ at z$\sim$4.5 to $\sim$ 1/3 Z$_{\sun}$ 
 at z$\sim$0.5. 
This has been validated with Zn, the more reliable metallicity inidicator by \citet{Kul07}.
 
To date, few observations have been made of $z < 1.5$ sub-DLAs due to the lack of spectrographs with enough sensitivity at short wavelengths and the 
the paucity of known sub-DLAs in this redshift range.
 Redshifts $z < 1.5$ span ~70$\%$ of the age of the Universe 
 (assuming a concordance cosmology of $\Omega_{m}=0.3, \Omega_{\Lambda}=0.7$).
 This redshift range is clearly important for understanding the nature of sub-DLA systems and galactic chemical evolution as well.
 
 Recently \citet{Per03a, Per06a, York06, Kul07, Pro06, Kh07, Mei07} have suggested that sub-DLAs may be 
 more metal rich than DLAs, and may contribute significantly to the cosmic metal budget. Several sub-DLA systems have actually been seen with super-solar abundances 
 \citep{Kh04, Per06a, Pro06, Mei07}. Super-solar abundances have also been reported in strong Lyman-$\alpha$ forest lines with log \nhI$<$16, although ionisation correction
 factors are important and not easily constrained \citep{Char03, Ding03, Mas05, Sch07}. 

As the DLA systems dominate the \nhI weighted mean metallicity, previous studies have been dominated by DLAs.
At least an equally large sample of sub-DLAs is needed to determine the overall metallicity evolution of QSO absorbers.
 The H I column density distribution for QSO absorbers shows that systems with  
 19 $\la$ log N$_{\rm H \ I} \la$ 20.3 are $\sim$8 times more numerous than the higher column density classical damped Ly$\alpha$ systems, 
 making sub-DLAs more readily available probes of neutral gas in the distant Universe.

 Many elements have been detected in QSO absorber systems, including C, N, O, Mg, Si, S, Ca, Ti, Cr, Mn, Fe, Ni, and Zn. Zn is 
 often the preferred tracer of the gas-phase metallicity as it is relatively undepleted in the Galactic ISM, especially 
 when the fraction of H in molecular form is low, as is the case in most DLAs. Zn also tracks the Fe abundance in Galactic stars (e.g., \citet{Niss04}), 
 and the lines of Zn II $\lambda$$\lambda$ 2026,2062 are relatively weak and typically unsaturated.
 These lines can be covered with ground based spectroscopy over a wide range of redshifts, 0.65 $\la$ z $\la$ 3.5, which covers a large portion ($\sim 45\%$) of the 
 history of the Universe. Abundances of refractory elements such as Cr and Fe relative to Zn also give a measure of the amount of dust extinction \citep{York06}.

The number of sub-DLAs studied to date is still small, with the largest samples coming from \citet{Per03a} and \citet{Mei07}.  
In this paper we add 10 new systems to our previous medium-resolution measurements of sub-DLAs taken with the Magellan Inamori Kyocera Echelle (MIKE) 
spectrograph  on the 6.5m Clay telescope at Las Campanas observatory. 
As the amount of published data on abundances of low redshift sub-DLAs is still small, we present these observations here with a more detailed analysis of the abundances (both
absolute and relative) and kinematics in a future paper with our complete sample.
The structure of this paper is as follows. 
In $\S$ 2, we discuss details of our observations and data reduction techniques. In $\S$ 3, column density determinations are discussed. $\S$ 4 gives information 
on the individual objects, and in $\S$ 5 we discuss the abundances of these absorbers. We also provide an appendix at the end of this paper, showing plots of the 
UV spectra with the fits to the Lyman-$\alpha$ lines.

\section{Observations and Data Reduction}

The observations presented here were made with the 6.5m Magellan-II Clay telescope and the Magellan Inamori Kyocera Echelle (MIKE) spectrograph \citep{Bern03} in 2007 March. 
This is a double sided spectrograph with both a blue and a red camera, providing for 
simultaneous wavelength coverage from $\sim$3340 \ang to $\sim$9400 \ang. Targets were observed in 
 multiple exposures of 1800 to 2700 sec each to minimize cosmic ray defects. The seeing was typically 
 $<$ 1$\arcsec$, averaging $\sim$ 0.7$\arcsec$. All of the target QSOs were observed 
 with the 1$\arcsec$x5$\arcsec$ slit and the spectra were binned 2x3 (spatial by spectral) during readout. 
The resolving power of the MIKE spectrograph is $\sim$19,000 and $\sim$25,000 on the red and blue sides respectively with a 1$\arcsec$ slit after binning. 
Table 1 gives a summary of the observations. 

These spectra were reduced using the MIKE pipeline reduction code in IDL developed by S. Burles, J. X. Prochaska, 
and R. Bernstein. Wavelengths were calibrated using a Th-Ar comparison lamp taken after each exposure. 
The data were first bias subtracted from the overscan region and flat-fielded. The data  were then sky-subtracted and the spectral orders were extracted
using the traces from flat field images. These extracted spectra were then corrected for heliocentric velocities and converted
to vacuum wavelengths. Each individual order was then extracted from the IDL structure created by the pipeline software and 
combined in IRAF using rejection parameters to reduce the effects of cosmic rays. These combined spectra were then normalized using a
polynomial function to fit the continuum. These functions were typically of order five or less. 

Our sample consists of 10 absorbers, 8 with 19.0 $<$ log \nhI $<$ 20.3 and 2 absorbers with log \nhI $<$ 19.0, at \za $\la$ 1.5, where few observations have been made to date. 
We selected these systems based on several criteria. Firstly, the absorbers were chosen to have 0.65 $\la$ \za $\la$ 1.5. Faint QSOs (m$_{V} \la$ 18.8) were avoided. 
All absorbers have known \nhI, with 19.0$\la$\nhI$<$20.3. 
Systems that were previously known to have extremely weak lines of Mg II (W$_{0}^{2796} \la$ 0.6 \ang), which are very unlikely to hold detectable Zn II lines at the thresholds available
to this program were also excluded. Systems with low \nhI (\nhI $\la$ 19.2) and faint QSOs (m$_{V}$ $\ga$ 18.5) were also avoided. 
Throughout this paper the QSO names are given in J2000 coordinates, except in table 1 where the original name, 
based on J1950 coordinates is given as the name \citep{HB87}.

We have attempted to be as unprejudiced as possible when selecting targets for observation. Of the $\sim$45 absorbers from \citet{Rao06} that are accessible with MIKE (i.e. had bright enough 
 background QSOs and were observable within the RA and declination constraints), we have observed 38 of these absorbers to date (\citet{Mei06, Mei07}, this work, and recent unpublished observations).
 Of the 7 that have  not been observed, one was not selected because of high (yet technically possible) declination, one was excluded due to very low absorption redshift which would make detection
 of the Zn II lines difficult at the wavelength cutoff of MIKE, and most of the others were skipped due to scheduling and observing constraints. Only two of these 7 absorbers have W$_{2796}<1.0$
 \ang, and none have  W$_{2796}<0.5$ \ang (i.e. the remaining systems are fairly strong Mg II systems), showing that this is a relatively unbiased sample. Furthermore, \citet{Kul07} showed 
 that systems with large  W$_{2796}$ have a wide range of [Zn/H], and that systems with high [Zn/H] have a large range of W$_{2796}$, or that selection by W$_{2796}$ in the low z \nhI surveys 
 does not appear to significantly bias the sample  towards higher metallicity systems.

\setlength{\tabcolsep}{4pt}
\begin{center} 
\begin{table*}
\begin{minipage}{180mm}
\caption{Summary of observed targets and their absorption systems.}

\begin{tabular}{ccccccccccc}

\hline
\hline
QSO		&	Alternate name		&m$_{\rm V \ or \ g}$	&	RA		&	Dec		&	z$_{em}$	&z$_{abs}$    	&log \nhI 	& E(B$-$V)$_{g-i}$$^{a}$& Exposure 		& Reference	\\
		&	or SDSS ID 		&			&			&			&			&		&		&			&	time		&		\\	
\hline	
J2000		&				& 		&	J2000		&	J2000		&			&      		&cm$^{-2}$   	&			& 	s 		&		\\
\hline
Q1037+0028	&	SDSS J103744.44+002809.2	&	18.4	&	10:37:44.40	&	+00:28:09.2	&	1.733		&1.4244	&	20.04$\pm$0.12 	&     	0.04		& 	10800		&	1	\\
Q1054-0020	&	SDSS J105440.98-002048.4	&	18.3	&	10:54:40.98	&	-00:20:48.4	&	1.021		&0.8301	&	18.95$\pm$0.18	&	$\cdots^{b}$ 	&	10800		& 	1	\\
$\cdots$	 	&	$\cdots$			&	$\cdots$   &	$\cdots$   	&	$\cdots$     	&	$\cdots$		&0.9514	&	19.28$\pm$0.02 	&	0.05		&	$\cdots$		& $\cdots$	\\
Q1215-0034	&	Q1213-002			&	17.0	&	12:15:49.81	&	-00:34:32.1	&	2.691		&1.5543	&	19.56$\pm$0.02 	&	0.02		&	5400		& 	2	\\
Q1220-0040	&	SDSS J122037.00-004032.4	&	18.5	&	12:20:37.00	&	-00:40:32.4	&	1.411		&0.9746	&	20.20$\pm$0.07 	&	-0.02		&	8100		& 	1	\\
Q1228+1018	&	Q1226+105			&	18.5 	&  	12:28:36.80	& 	+10:18:41.7	&    	2.305		&0.9376 	&	19.41$\pm$0.15 	&	0.00		&	7200		& 	3	\\
Q1330-2056	&	Q1327-206			&	17.0	&	13:30:07.70	&	-20:56:16.4	&	1.165		&0.8526	&	19.40$\pm$0.02 	&	N/A$^{c}$		&	3600		&  	4,5	\\
Q1436-0051	&	SDSS J143645.06-005150.6	&	18.5	&	14:36:45.03	&	 -00:51:50.6	&	1.275		&0.7377	&	20.08$\pm$0.11 	&	0.08 		&	5400		& 	1	\\
$\cdots$	 	&	$\cdots$			&   $\cdots$	&	$\cdots$		&	$\cdots$		&	$\cdots$		&0.9281	&	$<$18.8	     	&	$\cdots^{b}$	&	$\cdots$		& $\cdots$	\\
Q1455-0045	&	SDSS J145508.14-004507.4	&	18.0	&	14:55:08.14	&	-00:45:07.5	&	1.378		&1.0929	&	20.08$\pm$0.06 	&	0.02		&	8100		& 	1	\\
\hline
\end{tabular}
REFERENCES: (1) \citet{Nest04}; (2) \citet{Lanz87}; (3) \citet{Bart90}; (4) \citet{KB84}; (5) \citet{Berg87} \\
$^{a}$Determined using the colour excess $\Delta(g-i)=(g-i)-(g-i)_{med}$, where $(g-i)_{med}$ is the median QSO colour at the emission redshift from \citet{Rich01} and the relation 
$E(B-V)_{g-i}=\Delta(g-i)(1+z_{abs})^{-1.2}/1.506$ \citep{Kh04, York06}. \\
$^{b}$The extintinction is attributed to the system with the higher \nhI, although the true extincition could be due to either, or both systems.\\
$^{c}$This QSO was not observed in the SDSS
\end{minipage}
\end{table*}
\end{center}

\setlength{\tabcolsep}{6pt}
\begin{sidewaystable*}
\begin{minipage}{200mm}
\textbf{Table 2.} Rest-frame equivalent widths of key metal lines. Values and 1$\sigma$ errors are in units of m\ang.

\begin{tabular}{l c c c cccccccccc}        
\hline\hline                  
QSO		& 	z$_{abs}$	 &	 Mg I 	 	& 	Mg II		& 	Mg II	 	& 	Al II 	 	&	 Al III 	& 	Al III 		& 	Si II 		& 	Si II 		 & 	Ca II 		 &	 Ca II		&	  Cr II    	\\
		&			 &	 2852	 	& 	2796		& 	2803		& 	1670	 	&	 1854		& 	1862		& 	1526		& 	1808		 & 	3933		 &	 3969		&	 2056	\\
\hline 																							
 Q1037+0028	&	1.4244	 	&	 -$^{a}$ 		&	2443$\pm$19 	& 	2054$\pm$32 	& 	677$\pm$15 	&	169$\pm$12	& 	87$\pm$	8 	& 	746$\pm$49	& 	60$\pm$9 	 &	-		 &	-		&	$<$7		\\
 Q1054-0020	&	0.8301	 	&	 350$\pm$15 	& 	1180$\pm$14 	& 	1077$\pm$12 	&	-		&	415$\pm$18	& 	287$\pm$28	& 	-		& 	 	-	 & 	-		 &	 	-	&	 	$<$11	\\
  		&	0.9514	 	&	 131$\pm$22 	& 	769$\pm$19 	& 	542$\pm$27 	&	-		&	$<$22 		&	$<$11	 	& 	-		&	 $<$11		 & 	-	         &	 	-	&		$<$6	\\
 Q1215-0034	&	1.5543	 	&	 -	 	& 	1987$\pm$21 	& 	1543$\pm$26 	&	-		&	 	- 	& 	-	 	& 	-		&	 	-	 & 	-		 &	 	-	&	 	$<$10	\\
 Q1220-0040	&	0.9746	 	&	 152$\pm$20 	& 	1919$\pm$32 	& 	1688$\pm$32 	&	-		&	71$\pm$	22	& 	40$\pm$	22 	& 	-		& 	 	-	 & 	-		 &	 	-	&	 	$<$9	\\
 Q1228+1018	&	0.9376	 	&	 245$\pm$38 	& 	1678$\pm$40 	& 	1457$\pm$23 	&	-		&		- 	& 	-	 	& 	-		& 	 	-	 & 	-		 &	 	-	&	 	$<$6	\\
 Q1330-2056	&	0.8526	 	&	 293$\pm$66 	& 	2413$\pm$64 	& 	1807$\pm$41 	&	-		&	75$\pm$	24	& 	26$\pm$	19 	& 	-		& 	 	-	 & 	-		 &	 	-	&	 	$<$6	\\
 Q1436-0051	&	0.7377	 	&	 583$\pm$14 	& 	1113$\pm$10 	& 	1081$\pm$13 	&	-		&		- 	& 	-	 	& 	-		& 	 	-	 & 	396$\pm$15	 &	 	-	&	 	$<$20	\\
 		&	0.9281	 	&	 233$\pm$21 	& 	1105$\pm$40 	& 	867$\pm$33 	&	-		&	220$\pm$22 	& 	156$\pm$22 	& 	-		& 	46$\pm$	9	 & 	122$\pm$17	 &	 	-	&	 	$<$16	\\
 Q1455-0045	&	1.0929	 	&	 363$\pm$21 	& 	1659$\pm$10	& 	1586$\pm$11 	& 	804$\pm$14 	&	515$\pm$15 	& 	281$\pm$17 	& 	-		& 	 	-	 &	 		 &	       -	&	 	21 $\pm$ 7	\\
\hline\hline																					
 QSO		&	z$_{abs}$ 	&	 Mn II 		& 	Mn II 		& 	Mn II 	 	&	 Fe II 		&	Fe II 	 	& 	Fe II		&	 Fe II 		& 	Fe II 	 	 &	 	Fe II	 &	 Zn II 		&	 	Zn II	\\
 		&		 	&	 2576		& 	2594		& 	2606	 	&	 2260		&	2344	 	& 	2374		&	 2382		& 	2586		 &	 	2600	 &	 2026		&	 	2062	\\
\hline 																				
 Q1037+0028	&	1.4244	 	&	 75$\pm$20 	& 	82$\pm$18 	&	40$\pm$	13	& 	93$\pm$21 	& 	963$\pm$61 	& 	507$\pm$21 	& 	1442$\pm$55	 & 	839$\pm$51 	 &	1427$\pm$41	 &	$<$21		&	$<$15	\\
 Q1054-0020	&	0.8301	 	&	 42$\pm$11 	& 	39$\pm$8 		&	$<$8		&	$<$7		& 	586$\pm$7 	& 	277$\pm$9 	& 	805$\pm$6	 & 	548$\pm$6 	 &	831$\pm$6	 &	$<$10		&	$<$15	\\
 Q1054-0020	&	0.9514	 	&	$<$18		&	$<$10		&	$<$13		&	$<$6		& 	198$\pm$15 	& 	45$\pm$	9 	& 	420$\pm$18	 & 	208$\pm$27 	 &	453$\pm$27	 &	$<$9		&	$<$13	\\
 Q1215-0034	&	1.5543	 	&	$<$9		&	$<$8		&	$<$7		&	 24$\pm$5 	& 	507$\pm$14 	& 	267$\pm$9 	& 	861$\pm$20	 & 	449$\pm$10 	 &	896$\pm$24	 &	$<$8		&	$<$10	\\
 Q1220-0040	&	0.9746	 	&	$<$19		&	$<$22		&	$<$17		&	$<$8		& 	624$\pm$10 	& 	278$\pm$10 	& 	1049$\pm$32	 & 	566$\pm$25 	 &	932$\pm$22	 &	$<$9		&	$<$9	\\
 Q1228+1018	&	0.9376	 	&	$<$29		&	$<$20		&	$<$18		&	 64$\pm$12 	& 	692$\pm$12 	& 	417$\pm$22 	& 	989$\pm$9	 & 	660$\pm$19 	 &	1032$\pm$26	 &	$<$9		&	$<$8	\\
 Q1330-2056	&	0.8526	 	&	$<$9		&	$<$8		&	$<$8		&	$<$6		& 	271$\pm$51 	& 	58$\pm$	43 	& 	803$\pm$43	 & 	204$\pm$41 	 &	784$\pm$40	 &	$<$16		&	$<$6	\\
 Q1436-0051	&	0.7377	 	&	 207$\pm$17 	& 	134$\pm$14	&	85$\pm$15 	&	 116$\pm$24 	& 	757$\pm$17 	& 	580$\pm$16 	& 	847$\pm$9	 & 	785$\pm$11 	 &	926$\pm$16	 &	108$\pm$12	 &	101 $\pm$ 35	\\
 		&	0.9281	 	&	 43$\pm$16 	&	$<$22		&	102$\pm$32 	&	 25$\pm$7 	& 	341$\pm$16 	& 	189$\pm$13 	& 	501$\pm$13	 & 	307$\pm$36 	 &	471$\pm$23	 &	26$\pm$	7	 &	$<$21	\\
 Q1455-0045	&	1.0929	 	&	 29$\pm$6 	& 	50$\pm$14 	&	$<$19		&	$<$12		& 	902$\pm$11 	& 	470$\pm$22 	& 	1135$\pm$14	 & 	874$\pm$24 	 &	1226$\pm$21	 &	$<$15		&	$<$8	\\
																																											
\hline              
\end{tabular}
$^{a}$Blank entries indicate that the line was not measured due to lack of wavelength coverage, extremely low S/N at the wavelength extremes, or blends with telluric features. 
\end{minipage}
\setcounter{table}{2}

\end{sidewaystable*}
\setlength{\tabcolsep}{3pt}

\section{Determination of Column Densities}
    Column densities were determined from profile fitting with the package FITS6P \citep{Wel91}, which has
evolved from the code by \citet{VM77}. FITS6P iteratively minimizes the $\chi^{2}$ value between the data and
a theoretical Voigt profile that is convolved with the instrumental profile. 
The profile fit used multiple components, tailored to the individual system. For the central, core components, the effective Doppler parameters
($b_{eff}$) and radial velocities were determined from the weak and unsaturated lines, typically the Mg I $\lambda$ 2852 line. 
For the weaker components at higher radial velocities the $b_{eff}$ and component velocity values were determined from stronger transitions such as the 
Fe II $\lambda\lambda$ 2344, 2382 lines and the Mg II $\lambda\lambda$ 2796, 2803 lines. A set of $b_{eff}$ and $v$ values were thus
determined that reasonably fit all of the lines observed in the system. 
The same $b_{eff}$ values were used for all the species. The atomic data used in line identification and profile fitting are from \citet{Morton03}.

	 In general, if a multiplet was observed, the lines were fit simultaneously until convergence. 
For all of the systems, the Fe II $\lambda$ 2344, 2374, 2382 lines were fit simultaneously to determine a set of column
densities that fit the spectra reasonably well. Similarly, the Mg II $\lambda\lambda$ 2796, 2803 lines were also fit together. 
Significant saturation of the Mg II \lala 2796, 2803 and Al II $\lambda$ 1670 lines allowed for only lower limits to be placed on the column densities 
for these species. The Zn II $\lambda$ 2026.137 line is blended with a line of Mg I $\lambda$ 2026.477. The Mg I contribution to the line was estimated using the
Mg I $\lambda$ 2852 line, for which f$\lambda$ is $\sim$32 times that of the  Mg I $\lambda$ 2026 line. 
The Zn II components were then allowed to vary while the Mg I components were held fixed. 
N$_{\rm Cr \ II}$ was determined by simultaneously fitting the Cr II $\lambda$ 2056 line and the blended Cr II + Zn II $\lambda$ 2062 line, where the 
contribution from Zn II was estimated from the Zn II + Mg I $\lambda$ 2026 line. See also \citet{Kh04} and \citet{Mei07} for a discussion of the profile fitting scheme.
In this paper we adopt the standard notation:
	\begin{equation}\rm{[X/Y] = log(N_{\rm X}/N_{\rm H \ I}) - log (X/H)_{\sun}}\end{equation}
	Solar system abundances have been adopted from \citet{Lodd03}.

  We present the rest-frame equivalent widths ($W_{0}$) of the lines in Table 2. 
The 1$\sigma$ errors for the equivalent widths are given also and reflect both uncertainties in the 
continuum level and in the photon noise. If a certain line was not detected, the limiting equivalent width was determined from 
the local signal to noise ratio (S/N) and uncertainties in the continuum placement based on a 3-pixel element. 
Assuming a linear curve of growth, a corresponding 3$\sigma$ upper limit was also placed on the column density.

We give total column densities (the sum of the column densities in the individual components determined via profile fitting method) 
for these systems in table 3. Column densities were also obtained via the apparent optical depth (AOD)
method as a consistency check, and are also listed in table 3. For a discussion of the AOD method, see \citet{Sav96}. Errors for the AOD column densities include contributions 
from both the photon noise and continuum placement. In most cases, the column densities from the profile fitting and AOD methods 
agree to within the error bars, especially for the weak and unsaturated lines. 

\begin{sidewaystable*}
\begin{minipage}{220mm}
\textbf{Table 3.} Total column densities for the systems observed. Values are logarithmic, in units of cm$^{-2}$.
\end{minipage}
\centering                           
\begin{tabular}{lccccccccccccc}              
\hline\hline              
 QSO		& z$_{abs}$	 	& log \nhI		 	& 	Mg I		 	& 	Mg II		& 	Al II	 	& 	Al III		 	& 	Si II		 	& 	Ca II		 	&	 	Ti II	&	Cr II	 	&	 	Mn II		 &	 	Fe II		 	&	Zn II			\\
\hline         																								
 Q1037+0028	& 	1.4244	 	& 	20.04$\pm$0.12	 	& 	-		 	& 	$>$15.48 	& 	$>$14.15 	& 	13.19$\pm$0.02	 	& 	15.09$\pm$0.03	 	& 	-		 	&    -		 	&	$<$12.26	&	 12.53$\pm$0.05	 	&	 	14.84$\pm$0.02	 	&	$<$12.04		\\
 AOD		& 		 	& 			 	& 			 	& 	$>$14.44 	& 	$>$13.60 	&	13.15$\pm$0.03	 	& 	15.05$\pm$0.05	 	& 			 	&		 	&	 		&	 12.57$\pm$0.11	 	&	 	14.96$\pm$0.09	 	&				\\
 Q1054-0020A	& 	0.8301	 	& 	18.95$\pm$0.18	 	& 	12.55$\pm$0.01	 	& 	$>$14.83 	& 	-	 	& 	13.59$\pm$0.06	 	& 	-		 	& 	-		 	&    -		 	&	$<$12.47	&	 12.27$\pm$0.07	 	&	 	14.33$\pm$0.01	 	&	$<$11.76		\\
 AOD		& 		 	& 			 	& 	12.54$\pm$0.02	 	& 	$>$14.30 	& 		 	& 	13.65$\pm$0.04	 	& 			 	& 			 	&	 		&	 		&	 12.31$\pm$0.10	 	&	 	14.35$\pm$0.01	 	&				\\
 Q1054-0020B	& 	0.9514	 	& 	19.28$\pm$0.02	 	& 	12.04$\pm$0.03	 	& 	$>$13.72 	& 	-	 	&	$<$12.11		&	$<$14.25		& 	-		 	&	$<$12.36	&	$<$12.20	&	$<$11.92		&	 	13.66$\pm$0.01	 	&	$<$11.70		\\
 AOD		& 		 	& 			 	& 	12.04$\pm$0.07	 	& 	$>$13.59 	& 		 	& 			 	& 			 	& 			 	&		 	&	 		&	 			&	 	13.49$\pm$0.08	 	&				\\
 Q1215-0034	& 	1.5543	 	& 	19.56$\pm$0.02	 	& 	-		 	& 	$>$15.34 	& 	-	 	& 	-		 	& 	-		 	& 	-		 	&       -	 	&	$<$12.37	&	$<$11.63		&	 	14.39$\pm$0.01	 	&	$<$11.63		\\
 AOD		& 		 	& 			 	& 			 	& 	$>$14.25 	& 		 	& 			 	& 			 	& 			 	&		 	&	 		&	 		 	&	 	14.35$\pm$0.11	 	&				\\
 Q1220-0040	& 	0.9746	 	& 	20.20$\pm$0.07	 	& 	12.12$\pm$0.03	 	& 	$>$14.81 	& 	-	 	& 	12.62$\pm$0.06	 	& 	-		 	& 	-		 	&	$<$12.47	&	$<$12.38	&	$<$11.95		&	 	14.34$\pm$0.02	 	&	$<$11.69		\\
 AOD		& 		 	& 			 	& 	12.10$\pm$0.05	 	& 	$>$14.37 	&		 	& 	12.68$\pm$0.12	 	& 			 	& 			 	&		 	&	 	 	&	 		 	&	 	14.35$\pm$0.01	 	&				\\
 Q1228+1018	& 	0.9376	 	& 	19.41$\pm$0.02	 	& 	12.32$\pm$0.04	 	& 	$>$14.61 	&	-	 	& 	-		 	& 	-		 	& 	-		 	&	$<$11.65	&	$<$12.21	&	$<$12.14		&	 	14.58$\pm$0.01	 	&	$<$11.67		\\
 AOD		& 		 	& 			 	& 	12.32$\pm$0.05	 	& 	$>$14.36 	& 	 	 	& 			 	& 			 	& 			 	&		 	&	 	 	&	 		 	&	 	14.55$\pm$0.02	 	&				\\
 Q1330-2056	& 	0.8526	 	&	19.40$\pm$0.02	 	&	12.33$\pm$0.03	 	& 	$>$14.21 	&	-	 	& 	12.59$\pm$0.05	 	& 	-		 	& 	-		 	&	$<$11.45	&	$<$12.18	&	$<$11.61		&	 	13.80$\pm$0.01	 	&	$<$11.96		\\
 AOD		& 		 	& 			 	& 	12.37$\pm$0.09	 	& 	$>$14.12 	& 		 	& 	12.52$\pm$0.15	 	& 			 	& 			 	&		 	&	 	 	&	 		 	&	 	13.76$\pm$0.07	 	&				\\
 Q1436-0051A	& 	0.7377	 	&	20.08$\pm$0.11	 	& 	$>$12.92	 	& 	$>$14.42 	& 	-	 	& 		-	 	& 	-		 	& 	12.83$\pm$0.02	 	&	$<$11.71	&	$<$12.71	&	 13.01$\pm$0.02	 	&	 	14.94$\pm$0.02	 	& 12.67$\pm$0.05		\\																																																																																																																																																																																																																																				 	
 AOD		& 		 	& 			 	& 	$>$12.90	 	& 	$>$14.26 	& 		 	& 			 	& 			 	& 	12.79$\pm$0.01	 	&		 	&	 	 	&	 13.05$\pm$0.03	 	&	 	14.92$\pm$0.02	 	& 12.80$\pm$0.05		\\																																																																																																																																																																																																																																				 	
 Q1436-0051B	& 	0.9281	 	& 	$<$18.8		 	& 	12.49$\pm$0.04	 	& 	$>$15.12 	& 	-	 	& 	13.40$\pm$0.03	 	& 	15.02$\pm$0.05	 	& 	12.28$\pm$0.04	 	&	$<$12.71	&	$<$12.59	&	 12.45$\pm$0.05	 	&	 	14.20$\pm$0.01	 	& 12.29$\pm$0.06		\\																																																																																																																																																																																																																																				 	
 AOD		&		 	& 			 	& 	12.45$\pm$0.03	 	& 	$>$14.11 	& 		 	& 	13.31$\pm$0.03	 	& 	14.97$\pm$0.07	 	& 	12.15$\pm$0.05	 	&		 	&	 	 	&	 12.32$\pm$0.07	 	&	 	14.25$\pm$0.02	 	& 12.23$\pm$0.05		\\																																																																																																																																																																																																																																				 	
 Q1455-0045	&	1.0929	 	& 	20.08$\pm$0.06	 	& 	12.53$\pm$0.02	 	& 	$>$14.92 	& 	$>$13.83	& 	13.65$\pm$0.02	 	& 	14.64$\pm$0.10	 	& 	-		 	&	$<$12.44	& 	12.40$\pm$0.13 	&	 12.09$\pm$0.15	 	&	 	14.57$\pm$0.01	 	&	$<$11.91		\\
 AOD		&		 	& 			 	& 	12.51$\pm$0.02	 	& 	$>$14.45 	& 	$>$13.71 	& 	13.67$\pm$0.01	 	& 	14.84$\pm$0.12	 	& 			 	&		 	& 	12.75$\pm$0.17 	&	 12.20$\pm$0.10	 	&	 	14.61$\pm$0.02	 	&				\\
\hline         																											
\end{tabular}
\setcounter{table}{3}
\end{sidewaystable*}

\section{Discussion of Individual Objects}
\subsection{Q1037+0028 (z$_{em}$ = 1.733)}

\textbf{(System A:\za=1.4244):}
This QSO harbors a sub-DLA with log \nhI = 20.04$\pm$0.12 \citep{Rao06}. 
We detected absorption lines for Mg II, Al II, Al III, Si II, Mn II and Fe II in this sub-DLA system. 
The Mg I $\lambda$ 2852 line was blended with telluric features at $\sim$6916 \ang. 
A total of 11 components were used in the profile fitting for this system. No Zn II lines were detected in this system at S/N$\sim$16 in the region, 
and a limiting EW of W$_{0}<21$ m\ang. There were two small features that lined up with the components at 11 and 44 \kms from other lines, however, the equivalent width of the features
was significant at only a $\la$2$\sigma$ level. Total column densities for this and all other systems are given in Table 3.
The abundance based on Fe lines was [Fe/H]$=-$0.67$\pm$0.12. 
This system shows Al III/Al II $<-0.96$, a typical level of ionisation for a sub-DLA (e.g., \citet{Des03, Mei07}). 
Rest frame equivalent widths for the lines in this system are found in table 2, 
along with the results of the profile fitting in table 4. 
Velocity plots of several lines of interest are shown in Figure 1. 

There also appears to be a weaker system at \za=0.6572, with
W$_{0}$(Mg II 2796)=354 m\ang. We detect the Fe II $\lambda\lambda$ 2344, 2382, 2586, and 2600 lines along with the Mg I $\lambda$ 2852 line as well. 
The Fe II $\lambda$ 2374 line is blended with the Galactic Ca II $\lambda$ 3934 line for this system. \nhI cannot be determined for this weaker system due to the 
break in the continuum from the sub-DLA system at \za=1.4244 beyond the Lyman limit, so it is not included in our sample (which depends on known \nhI, as noted earlier).

\begin{minipage}{115mm}
\end{minipage}


\begin{table*}
\begin{minipage}{135mm}
\begin{center}
\caption{Column densities for the $z=$1.4244 absorber in Q1037+0028 with \nhI=20.04$\pm$0.12. Velocities and b$_{eff}$ values are in units of \kms, and column
  	 densities are in units of cm$^{-2}$. Errors in the column densities are the 1$\sigma$ values.}
\begin{tabular}{cccccccc}
\hline
\hline
v	&	b$_{eff}$	&		Mg II	&		Al II				&	Al III		 	&	Si II					&	Mn II			&	Fe II 		\\
\hline
$-$158	&	6.9		&	$>$	3.14E13	&	$>$	2.34E12		 		&	(6.59$\pm$1.40)E11	&	(4.26$\pm$0.59)E13			&	-			&	(1.63$\pm$0.13)E13		\\
$-$118	&	5.9		&	$>$	8.37E13	&	$>$	2.16E12		 		&	-			&	(5.69$\pm$0.98)E13			&	-			&	(3.04$\pm$0.21)E13		\\
$-$91	&	6.7		&	$>$	3.91E12	&	$>$	4.06E11		 		&	-			&	(6.52$\pm$0.20)E12			&	-			&	(2.92$\pm$0.76)E12		\\
$-$57	&	13.5		&	$>$	1.65E13	&	$>$	1.28E12		 		&	-			&	(3.65$\pm$0.46)E13			&	-			&	(1.65$\pm$0.14)E13		\\
$-$40	&	6.9		&	$>$	9.77E11	&	$>$	1.47E12		 		&	-			&	(2.98$\pm$0.60)E13			&	-			&	(1.69$\pm$0.17)E13		\\
$-$19	&	10.0		&	$>$	6.90E13	&	$>$	4.60E12		 		&	(1.37$\pm$0.17)E12	&	(7.21$\pm$0.93)E13			&	-			&	(3.92$\pm$0.21)E13		\\
11	&	8.1		&	$>$	1.23E14	&	$>$	5.88E13		 		&	(1.12$\pm$0.05)E13	&	(7.10$\pm$0.61)E14			&	(2.07$\pm$0.29)E12	&	(3.08$\pm$0.25)E14		\\
44	&	7.9		&	$>$	2.69E15	&	$>$	6.80E13		 		&	(2.42$\pm$0.18)E12	&	(2.53$\pm$0.48)E14			&	(1.33$\pm$0.26)E12	&	(2.36$\pm$0.17)E14		\\
100	&	11.2		&	$>$	7.16E12	&	 	(4.64$\pm$0.98)E11		&	-			&	(6.79$\pm$2.23)E12			&	-			&	(5.65$\pm$0.89)E12		\\
147	&	12.6		&	$>$	1.56E13	&		(9.89$\pm$1.16)E11		&	-			&	(2.43$\pm$0.32)E13			&	-			&	(1.33$\pm$0.11)E13		\\
183	&	4.6		&	$>$	4.57E12	&		-		 		&	-			&	-					&	-			&	-		\\
\hline
\end{tabular}
\end{center}
\end{minipage}
\end{table*}

\subsection{Q1054-0020 (z$_{em}$=1.021)}
\textbf{(System A:\za=0.8301):}
System A is a strong Lyman-limit system  with \nhI=18.95$\pm$0.18 \citep{Rao06}. 
Five components were used in the profile fits for system A. Lines of Mg I , Mg II, Mn II, and Fe II were all detected. No Zn II lines were seen at S/N$\sim$23 in the region, 
We did not detect Zn II $\lambda$ 2026 with a limiting EW of \w$<$ 10 m\ang, and [Zn/H]$<+$0.18. This system has a ratio of N$_{\rm Fe \ II}$/N$_{\rm H \ I}$$=-$0.09$\pm$0.18 dex. 
The true ratio may be lower due to ionisation (see our later comments on ionisation corrections at low \nhI). 
We also detected lines of Mn II \lala 2576, 2594 with [Mn/Fe]$=-$0.09$\pm$0.07.
The Al II $\lambda$ 1670  line lies below the wavelengths covered by MIKE, so the Al III/Al II ratio could not be measured in this system. We did however detect strong Al III \lala 1854, 162 lines. 
Velocity plots of several lines of interest are shown in Figure 2. Rest frame equivalent widths for the lines in this system can be found in table 2, 
with the results of the profile fitting analysis in table 5.

\textbf{(System B:\za=0.9514):} A total of 7 components were used to fit system B with \nhI=19.28$\pm$0.02  \citep{Rao06}. Only lines of Mg I, Mg II, and Fe II were detected in this system. We were able to constrain the metallicity of 
this system based on the non detection of the Zn II $\lambda$ 2026 line to be [Zn/H]$<-$0.21 with S/N$\sim$30 in the region. This system also has a low metallicity based on Fe with
[Fe/H]$=-$1.09$\pm$0.02, although dust depletion may be important. 
The Al II $\lambda$ 1670 line lies below the covered wavelengths. The Al III \lala 1854, 1862 lines were covered but not detected with \w$<$22 m\ang. 
Velocity plots of several lines of interest are shown in Figure 3. The results of the profile fitting analysis are found in table 6,
and the rest frame equivalent widths for the lines in this system in table 2.

\begin{table*}
\begin{minipage}{115mm}
\begin{center}
\caption{ Same as table 3, but for the \za=0.8301 system in Q1054-0020 with \nhI=18.95$\pm$0.18.}
\begin{tabular}{ccccccc}

\hline
\hline
v	&b$_{eff}$	&	Mg I			&	Mg II			&	Al III			&	Mn II			&	Fe  II			\\
\hline
$-$25	&	9.1	&	(4.53$\pm$0.41)E11	&	$>$1.07E14		&	(9.41$\pm$1.86)E12	&	-			&	(3.07$\pm$0.08)E13	 	\\
2	&	10.3	&	(1.53$\pm$0.08)E12	&	$>$2.00E14		&	(1.80$\pm$0.43)E13	&	(8.30$\pm$1.68)E11	&	(7.96$\pm$0.20)E13	 	\\
19	&	8.1	&	(6.91$\pm$0.58)E11	&	$>$ 8.82E12		&		-		&	(4.12$\pm$1.50)E11	&	(2.63$\pm$0.10)E13	 	\\
49	&	13.4	&	(8.89$\pm$0.51)E11	&	$>$ 3.59E14		&	(1.15$\pm$0.18)E13	&	(6.09$\pm$1.73)E11	&	(7.94$\pm$0.15)E13	 	\\
104	&	15.8	&	-			&	(1.23$\pm$0.12)E12	&	        -		&	-			&	-	 			\\
\hline
\end{tabular}
\end{center}
\end{minipage}
\end{table*}

\begin{table*}
\begin{minipage}{70mm}
\begin{center}
\caption{ Same as table 3, but for the \za=0.9514 system in Q1054-0020 with \nhI=19.28$\pm$0.02.}
\begin{tabular}{ccccc}
\hline
\hline
v	&b$_{eff}$	&	Mg I			&	Mg II 			&	Fe II			\\
\hline
$-$447	&	8.0	&	-			&	(1.15$\pm$0.11)E12	&	-			\\
$-$90	&	5.1	&	(1.71$\pm$0.28)E11	&	$>$6.74E12		&	(4.86$\pm$0.27)E12	\\
$-$71	&	4.6	&	-			&	$>$1.84E12		&	-			\\
$-$13	&	10.0	&	(2.23$\pm$0.31)E11	&	$>$1.24E13		&	(9.84$\pm$0.34)E12	\\
49	&	3.5	&	(7.26$\pm$2.45)E10	&	$>$1.43E12		&	(1.71$\pm$0.23)E12	\\
72	&	8.0	&	(5.24$\pm$0.41)E11	&	$>$2.64E13		&	(2.66$\pm$0.06)E13	\\
87	&	10.3	&	(1.07$\pm$0.33)E11	&	$>$3.07E12		&	(2.69$\pm$0.33)E12	\\
\hline
\end{tabular}
\end{center}
\end{minipage}
\end{table*}

\begin{minipage}{115mm}
\end{minipage}

\begin{minipage}{115mm}
\end{minipage}

\subsection{Q1215-0034 (z$_{em}$=2.691)} 
\textbf{(System A:\za=1.5543):}
This fairly high redshift QSO harbors a sub-DLA with log \nhI=19.56$\pm$0.02 \citep{Rao06}. 
Fourteen components were used in the profile fit for this system, spanning $\sim$390 \kms in velocity space. 
Several of the components were detected only in the Mg II $\lambda\lambda$ 2796, 2803 lines. We detected lines from Mg II $\lambda\lambda$ 2796, 2803 and Fe II 
\lala  2344, 2374, 2382, 2586, and 2600. 
The Al II $\lambda$ 1670 and Al III \lala 1854, 1862 lines were covered, but are blended with features inside the Lyman-$\alpha$ forest. 
The Mg I $\lambda$ 2852 line was blended with telluric features at $\sim$7284 \ang. We did not detect Zn II at S/N $\sim$30 in the region, but 
we could place an upper limit of [Zn/H]$<-$0.56, and based on the Fe II lines an abundance of [Fe/H]$=-$0.64$\pm$0.02. Velocity plots of lines of interest are shown in Figure 4.
The results of the profile fitting analysis are given in table 7, with the rest frame equivalent widths for the lines in this system in table 2.

\begin{table*}
\begin{minipage}{60mm}
\begin{center}
\caption{ Same as table 3, but for the \za=1.5543 system in  Q1215-0034 with \nhI=19.56$\pm$0.02.}
\begin{tabular}{cccc}
\hline
\hline
 v 	&b$_{eff}$	&	Mg II			&		Fe II			\\
\hline
$-$271	&	8.0	&	(1.42$\pm$0.09)E12	&		-				\\
$-$217	&	6.7	&	(1.84$\pm$0.10)E12	&		-				\\
$-$188	&	9.7	&	$>$7.931E12		&		(2.52$\pm$0.41)E12	\\
$-$154	&	9.0	&	$>$1.10E13		&		(2.28$\pm$0.40)E12	\\
$-$131	&	7.4	&	$>$1.04E12		&		-		\\
$-$103	&	13.9	&	$>$1.77E12		&		-		\\
$-$75	&	8.7	&	$>$5.48E12		&		(5.22$\pm$0.44)E12	\\
$-$45	&	10.0	&	$>$3.30E13		&		(1.75$\pm$0.06)E13	\\
$-$3	&	9.7	&	$>$2.06E15		&		(9.13$\pm$0.36)E13	\\
14	&	15.5	&	$>$6.78E13		&		(1.18$\pm$0.03)E14	\\
42	&	10.9	&	$>$1.57E13		&		(6.80$\pm$0.61)E12	\\
70	&	10.9	&	(2.92$\pm$0.13)E12	&		-			\\
102	&	5.4	&	(1.16$\pm$0.09)E12	&		-			\\
120	&	6.2	&	(1.87$\pm$0.10)E12	&		(1.77$\pm$0.37)E12	\\
\hline
\end{tabular}
\end{center}
\end{minipage}
\end{table*}

\begin{minipage}{115mm}
\end{minipage}

\subsection{Q1220-0040 (z$_{em}$=1.411)}
\textbf{(System A:\za=0.9746):}
There is a sub-DLA with log \nhI=20.20$\pm$0.07 in the UV spectrum of this QSO \citep{Rao06}. We detect lines of Al III, Mg I, Mg II, and Fe II in this sub-DLA. 
A total of 9 components were used to fit the line profiles for this system.
The Al II $\lambda$ 1670 line lines below the wavelengths accessible to MIKE, so we were unable to determine the Al III/Al II ratio for this system.
Al III was detected only in the components at v$=-$197 and v$=-$162 \kms, and not in the components where other ions were the bulk of the absorption from other ions
was seen, such as the v=2, v=38, or v=74 \kms components. 
We did not detect any Zn II lines in this system at S/N$\sim$23 in the region, and place an upper limit on the Zn metallicity of [Zn/H]$<-$1.14. The Fe II abundance is similarly low
for this system, with [Fe/H]$=-$1.33$\pm$0.07. We show velocity plots of several lines in Figure 5. The results of the profile fitting analysis are found in table 8, with the 
rest frame equivalent widths for this system in table 2.

\begin{table*}
\begin{minipage}{80mm}
\begin{center}
\caption{ Same as table 3, but for the \za=0.9746 system in Q1220-0040 with \nhI=20.20$\pm$0.07.}
\begin{tabular}{cccccc}
\hline
\hline
v	&b$_{eff}$	&	Al III			&	Mg I			&	Mg II			&	Fe II			\\
\hline	
$-$197	&	4.9	&	(1.50$\pm$0.33)E12	&	-			&	$>$6.51E12		&	(1.03$\pm$0.27)E12	\\
$-$162	&	5.6	&	(2.70$\pm$0.42)E12	&	-			&	$>$1.53E13		&	-			\\
$-$135	&	11.1	&	-			&	-			&	$>$4.45E12		&	-			\\
$-$100	&	4.6	&	-			&	-			&	$>$4.19E12		&	(2.08$\pm$0.29)E12	\\
$-$45	&	8.0	&	-			&	-			&	$>$3.10E12		&	(1.52$\pm$0.32)E12	\\
2	&	14.6	&	-			&	(6.71$\pm$0.52)E11	&	$>$4.95E14		&	(1.22$\pm$0.02)E14	\\
38	&	15.8	&	-			&	(3.91$\pm$0.48)E11	&	$>$5.58E13		&	(4.91$\pm$0.11)E13	\\
75	&	15.2	&	-			&	(2.59$\pm$0.44)E11	&	$>$5.86E13		&	(4.08$\pm$0.09)E13	\\
106	&	9.0	&	-			&	-			&	$>$6.07E12		&	(2.60$\pm$0.36)E12	\\
\hline
\end{tabular}
\end{center}
\end{minipage}
\end{table*}

\begin{minipage}{115mm}
\end{minipage}

\subsection{Q1228+1018 (z$_{em}$=2.305)}
\textbf{(System A:\za=0.9376):}
This moderately high redshift QSO has a sub-DLA in its spectrum with log \nhI=19.41$\pm$0.02 \citep{Rao06}. We detected strong lines of Mg I, Mg II, and Fe II in this sub-DLA. A total of 
9 componenets were used in the Voigt profile fits. Due to the high redshift of the background QSO, several lines were blended with Lyman-$\alpha$ forest clouds, such as 
Al III \lala 1854, 1862 and Si II $\lambda$ 1808. The Zn II \lala 2026, 2062 lines also fell inside the Lyman-$\alpha$ forest, but lay between strong absorption features, especially the
Zn II $\lambda$ 2026 line. There is a strong forest line in the blue end of the  expected position of the Zn II feature, 
as can be seen in the velocity plots in Figure 6, however the components which are most likely to 
hold Zn II are at v=2, v=26, and v=51 \kms, which are unblended. Nonetheless, no Zn II is seen, and we place an upper limit on the Zn metallicity of [Zn/H]$<-$0.37
 with S/N$\sim$26 in the region. The Fe abundance for this sub-DLA is high compared to typical DLA Fe abundances, with [Fe/H]$=-$0.30$\pm$0.02. 
The results of the profile fitting analysis are given in table 9, with the rest frame equivalent widths for the lines in this system given in table 2. 
 
\begin{table*}
\begin{minipage}{80mm}
\begin{center}
\caption{ Same as table 3, but for the \za=0.9376 system in Q1228+1018 with \nhI=19.41$\pm$0.02.}
\begin{tabular}{ccccc}
\hline
\hline
v	&b$_{eff}$	&	Mg I			&	Mg II			&	Fe II		\\
\hline
$-$44	&	20.2	&		-		&	(3.05$\pm$0.27)E12	&	(3.05$\pm$0.58)E12	\\
$-$15	&	8.9	&	(2.16$\pm$0.69)E11	&	$>$6.96E13		&	(1.86$\pm$0.08)E13	\\
2	&	9.2	&	(1.18$\pm$0.66)E11	&	$>$2.46E13		&	(3.97$\pm$0.14)E13	\\
26	&	9.9	&	(9.72$\pm$1.02)E11	&	$>$1.64E14		&	(2.29$\pm$0.05)E14	\\
51	&	11.1	&	(4.93$\pm$0.78)E11	&	$>$6.42E13		&	(3.95$\pm$0.12)E13	\\
77	&	8.0	&	(1.88$\pm$0.61)E11	&	$>$5.86E13		&	(4.31$\pm$0.13)E12	\\
109	&	7.4	&		-		&	$>$1.39E13		&	(6.54$\pm$0.45)E12	\\
126	&	4.9	&	(1.04$\pm$0.36)E11	&	$>$8.37E12		&	(1.72$\pm$0.38)E12	\\
155	&	14.9	&		-		&	(4.17$\pm$0.25)E12	&	(1.92$\pm$0.48)E12	\\
\hline
\end{tabular}
\end{center}
\end{minipage}
\end{table*}

\begin{minipage}{115mm}
\end{minipage}

\subsection{Q1330-2056 (z$_{em}$=1.165)}
\textbf{(System A:\za=0.8526):}
In the spectrum of this relatively bright (m$_{V}=17.0$) QSO, there is a sub-DLA at \za = 0.8526 with log \nhI=19.40$\pm$0.02 \citep{Rao06}. We detected strong absorption from 
Al III, Mg I, Mg II, and Fe II. The Al II $\lambda$ 1670 line was below the covered wavelengths, so the Al III/Al II ratio could not be determined for this system. 
This system required 15 components in the Voigt 
profile models for an adequate fit. The components spanned $\sim$650 \kms in velocity space. 
The Zn II $\lambda$ 2026 line fell near a bad column in the CCD on the blue side. 
However, the components where Zn II is most likely, such as the components with the strongest Mg I absorption, are outside of the affected area. We have highlighted the affected area 
in grey in the plot of the Zn II $\lambda$ 2026 line in Figure 7.
No Zn II was detected, and an upper limit of 
[Zn/H]$<-$0.07 was placed on the system at S/N$\sim$18 in the region. The Fe abundance was low, with [Fe/H]$<-$1.07$\pm$0.02. We give velocity plots of several 
lines in Figure 7. The results of the profile fitting analysis are shown in table 10, with the rest frame equivalent widths for the lines in this system given in table 2.

\begin{table*}
\begin{minipage}{90mm}
\begin{center}
\caption{ Same as table 3, but for the \za=0.8526 system in Q1330-2056 with \nhI=19.40$\pm$0.02.}
\begin{tabular}{cccccc}
\hline
\hline
v	&b$_{eff}$	&	Al III			&	Mg I		&	Mg II			&	Fe II 			\\
\hline
$-$439	&	6.2	&	(6.39$\pm$1.74)E11	&	-			&	$>$9.07E12		&	(2.23$\pm$0.37)E12	\\
$-$365	&	9.9	&	(1.91$\pm$0.23)E12	&	(1.80$\pm$0.40)E11	&	$>$3.09E13		&	(8.13$\pm$0.50)E12	\\
$-$295	&	4.3	&	-			&	-			&	(1.44$\pm$0.28)E12	&	-			\\
$-$262	&	6.4	&	-			&	(1.03$\pm$0.35)E11	&	(2.23$\pm$0.30)E12	&	-			\\
$-$192	&	9.7	&	(6.60$\pm$1.96)E11	&	(2.84$\pm$4.28)E11	&	$>$2.05E13		&	(6.50$\pm$0.47)E12	\\
$-$121	&	7.2	&	-			&	-			&	(2.28$\pm$0.33)E12	&	-			\\
$-$101	&	10.4	&	-			&	(2.07$\pm$0.43)E11	&	$>$5.25E12		&	(1.26$\pm$0.42)E12	\\
$-$74	&	12.8	&	-			&	-			&	$>$3.49E12		&	(1.56$\pm$0.46)E12	\\
$-$10	&	10.9	&	-			&	-			&	(2.31$\pm$0.31)E12	&	-			\\
21	&	9.2	&	-			&	(2.97$\pm$0.43)E11	&	$>$2.91E13		&	(9.65$\pm$0.52)E12	\\
46	&	10.4	&	-			&	(1.17$\pm$0.41)E11	&	$>$3.83E12		&	-			\\
67	&	11.4	&	-			&	(1.09$\pm$0.41)E11	&	$>$6.20E12		&	(3.02$\pm$0.47)E12	\\
99	&	7.5	&	-			&	(4.30$\pm$0.48)E11	&	$>$2.74E13		&	(1.75$\pm$0.07)E13	\\
119	&	7.9	&	-			&	(1.58$\pm$0.41)E11	&	$>$7.84E12		&	(7.74$\pm$0.50)E12	\\
136	&	6.4	&	(6.36$\pm$1.78)E11	&	(1.12$\pm$0.38)E11	&	$>$1.04E13		&	(4.31$\pm$0.42)E12	\\
\hline
\end{tabular}
\end{center}
\end{minipage}
\end{table*}

\begin{minipage}{115mm}
\end{minipage}

  \begin{figure*}
  \begin{minipage}{160mm}
  \begin{center}
 $\begin{array}{c@{\hspace{0.0in}}c}
  \multicolumn{1}{l}{\mbox{\bf }} &
  	\multicolumn{1}{l}{\mbox{\bf }} \\ [0.0cm]
  		\includegraphics[angle=90, width=5.5in]{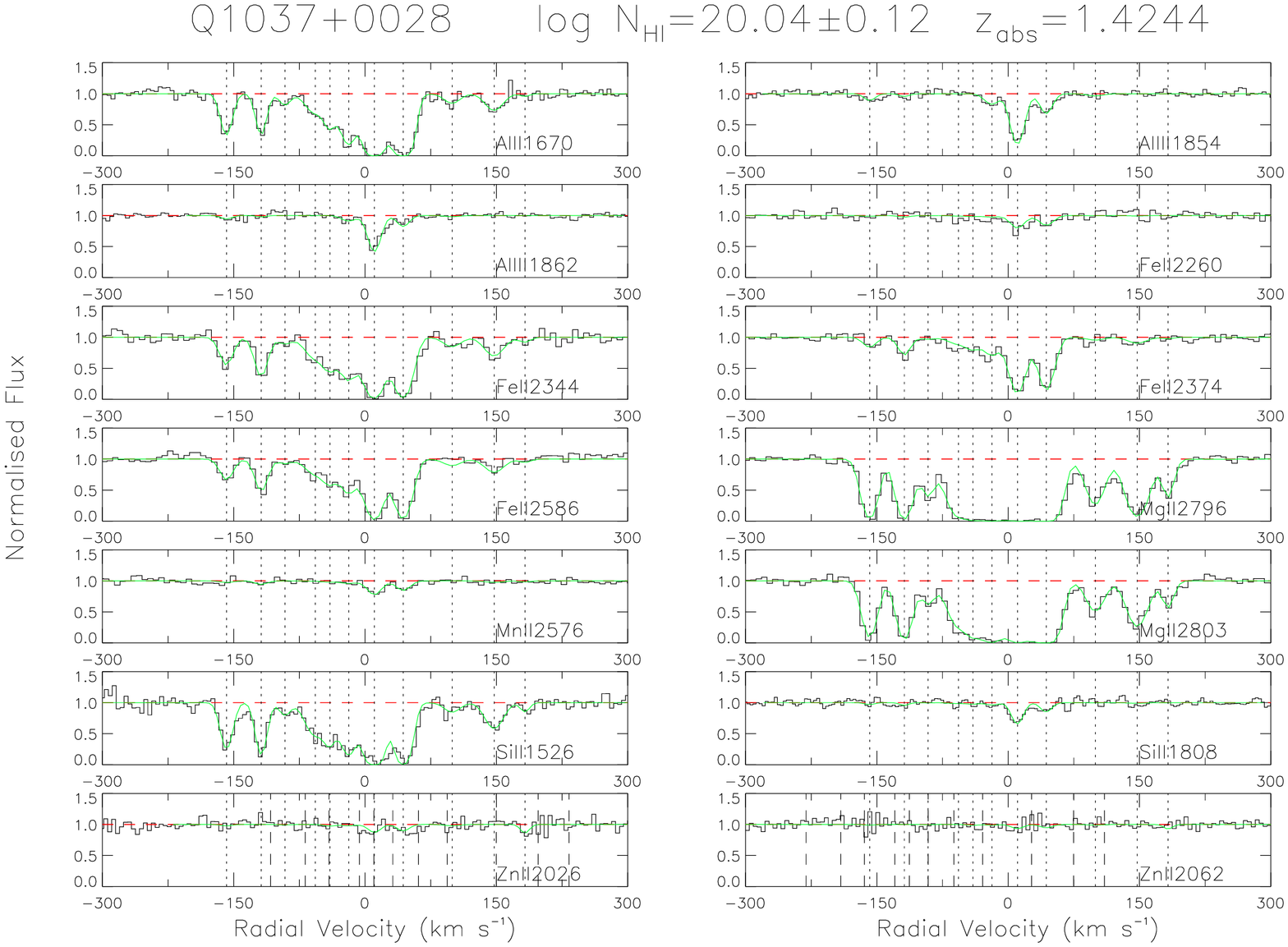} \\ [0.5cm] \includegraphics[angle=90, width=5.5in]{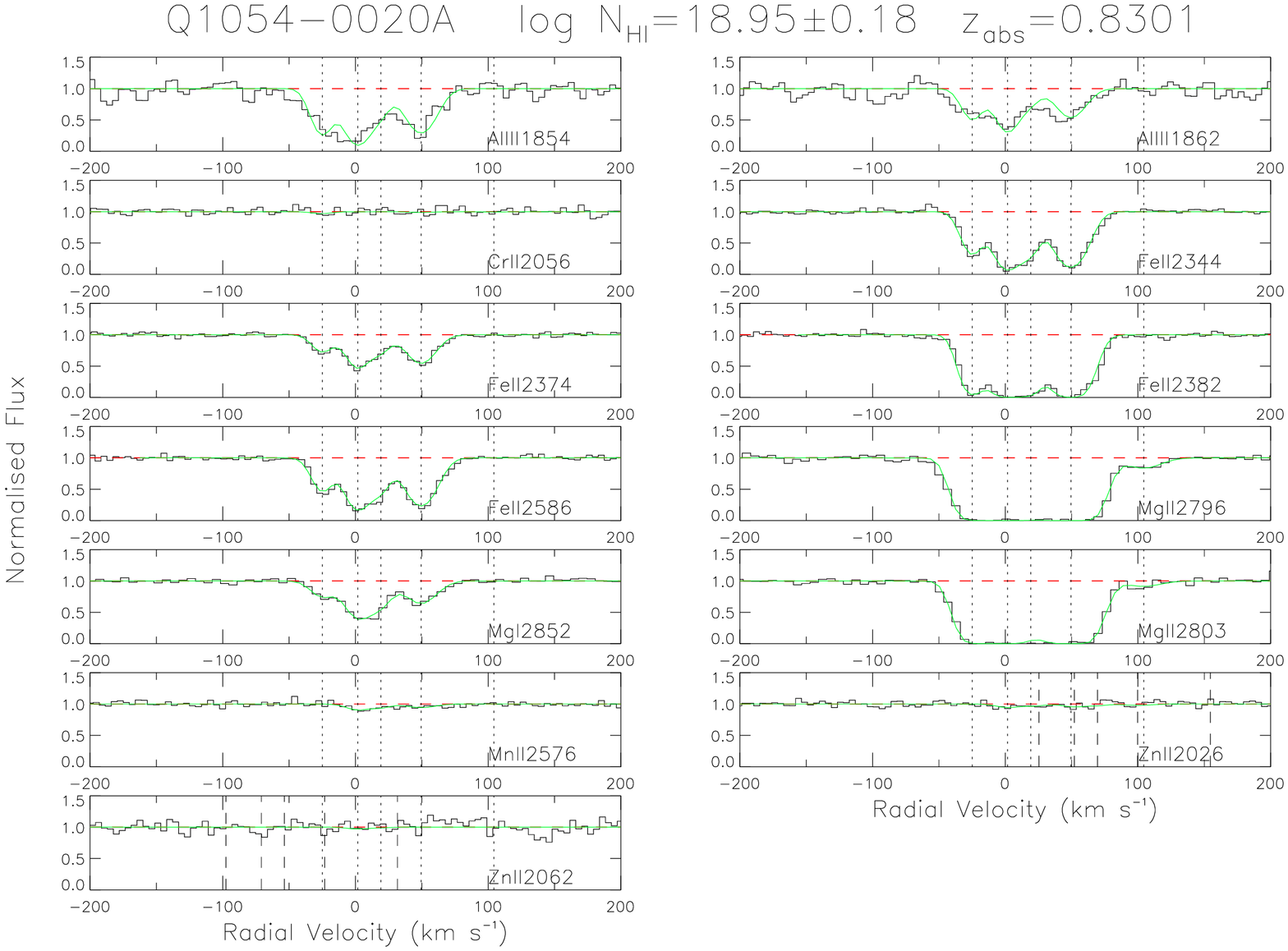} \\ [0.0cm]

  \end{array}$
  \end{center}
 \small{\textbf{ Figures 1 and 2:} Velocity plots for several lines of interest in the $z=$1.4244 system (top) in the spectrum of Q1037+0028
 and the \za=0.8301 system in Q1054-0020 (system A, bottom). The solid green line indicates the theoretical profile fit to the spectrum, and the dashed red line is the continuum level.
 The vertical dotted lines indicate the positions of the components that were used in the fit. In the cases of the Zn II $\lambda\lambda$ 2026,2062
 lines, which have other lines nearby, the long dashed vertical lines indicate the positions of the components for Mg I (former case), and Cr II (latter case). }
  \end{minipage}
  \end{figure*}

\begin{figure*}
  \begin{minipage}{160mm}
  \begin{center}
 $\begin{array}{c@{\hspace{0.0in}}c}
  \multicolumn{1}{l}{\mbox{\bf }} &
  	\multicolumn{1}{l}{\mbox{\bf }} \\ [0.0cm]
  		\includegraphics[angle=90, width=5.5in]{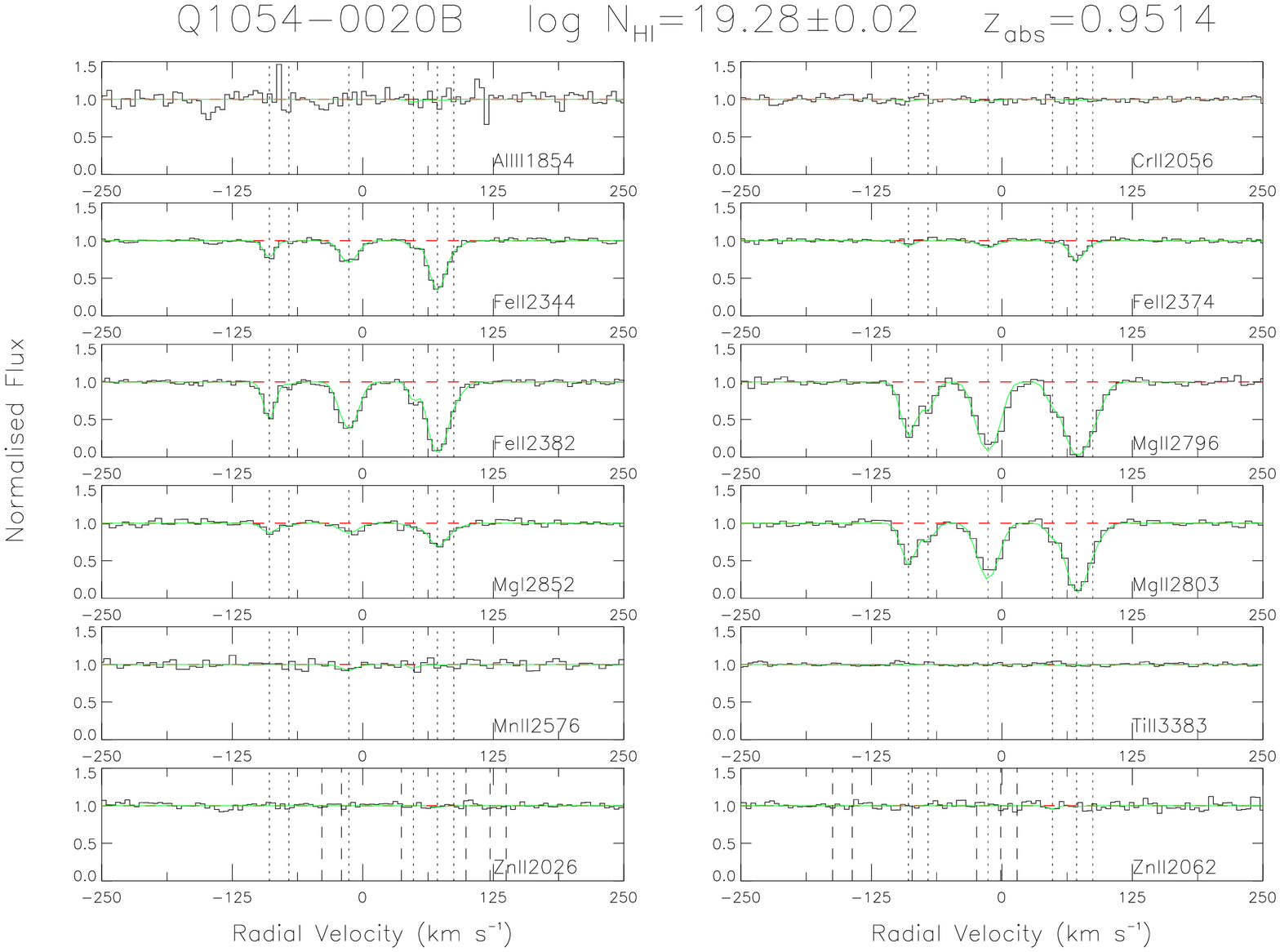} \\ [0.5cm] \includegraphics[angle=90, width=5.5in]{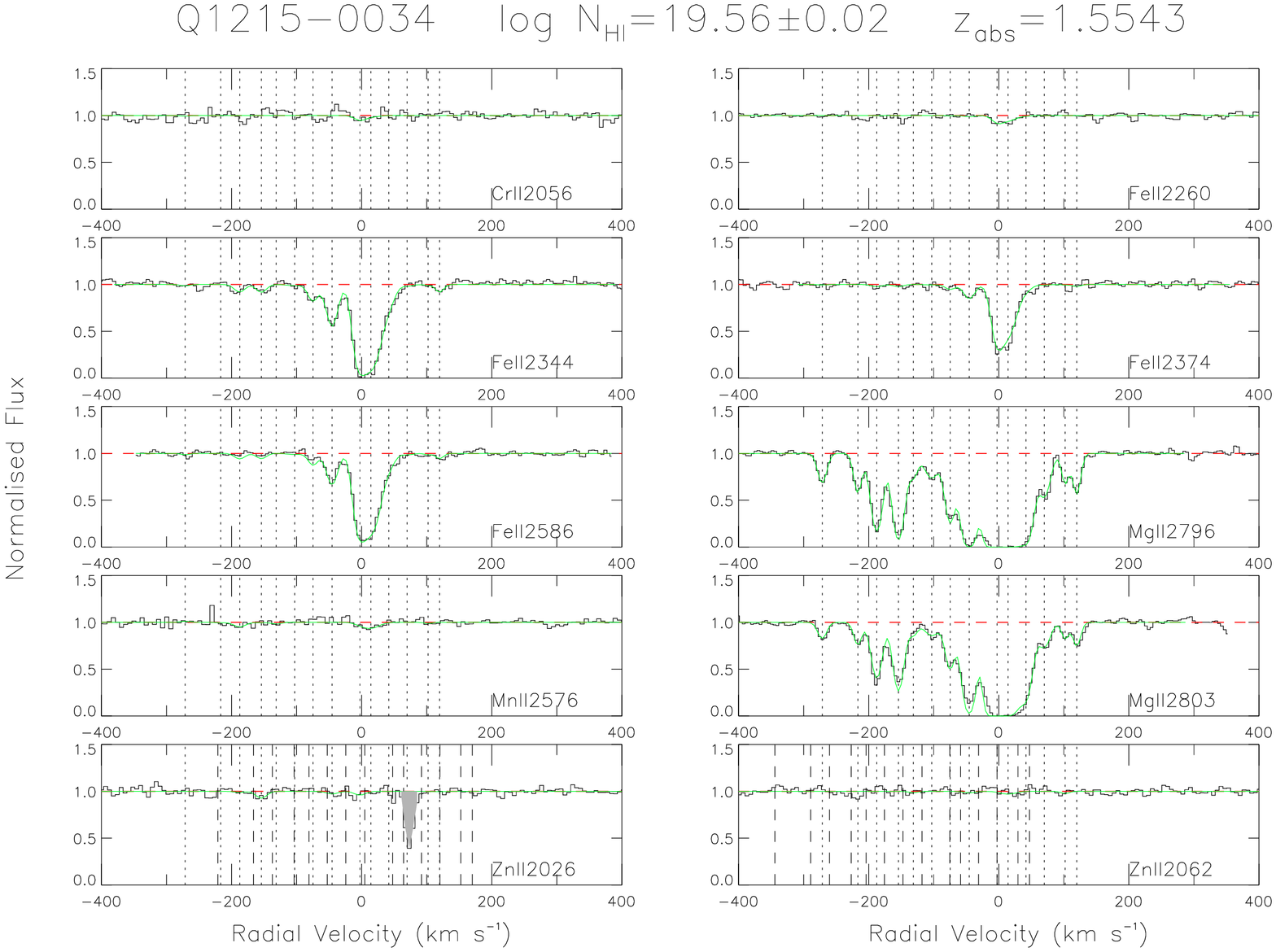} \\ [0.0cm]

  \end{array}$
  \end{center}
  \small{\textbf{ Figures 3 and 4:} Same as Figure 1, but for the \za=0.9514 system in Q1054-0020 (system B, top), and the \za=1.5543 system in Q1215-0034 (bottom). 
  For the \za=1.5543 system in Q1215-0034, the area shaded in grey in the panel displaying the Zn II $\lambda$ 2026 line is an unidentified feature, not due to absorption from
  Zn II. This may be due to Mg I 2026, however, the Mg I 2852 line was blended with telluric features, and we could therefore not estimate the Mg I contribution to this line.}
  \end{minipage}
  \end{figure*}

\begin{figure*}
  \begin{minipage}{160mm}
  \begin{center}
 $\begin{array}{c@{\hspace{0.0in}}c}
  \multicolumn{1}{l}{\mbox{\bf }} &
  	\multicolumn{1}{l}{\mbox{\bf }} \\ [0.0cm]
  		\includegraphics[angle=90, width=5.5in]{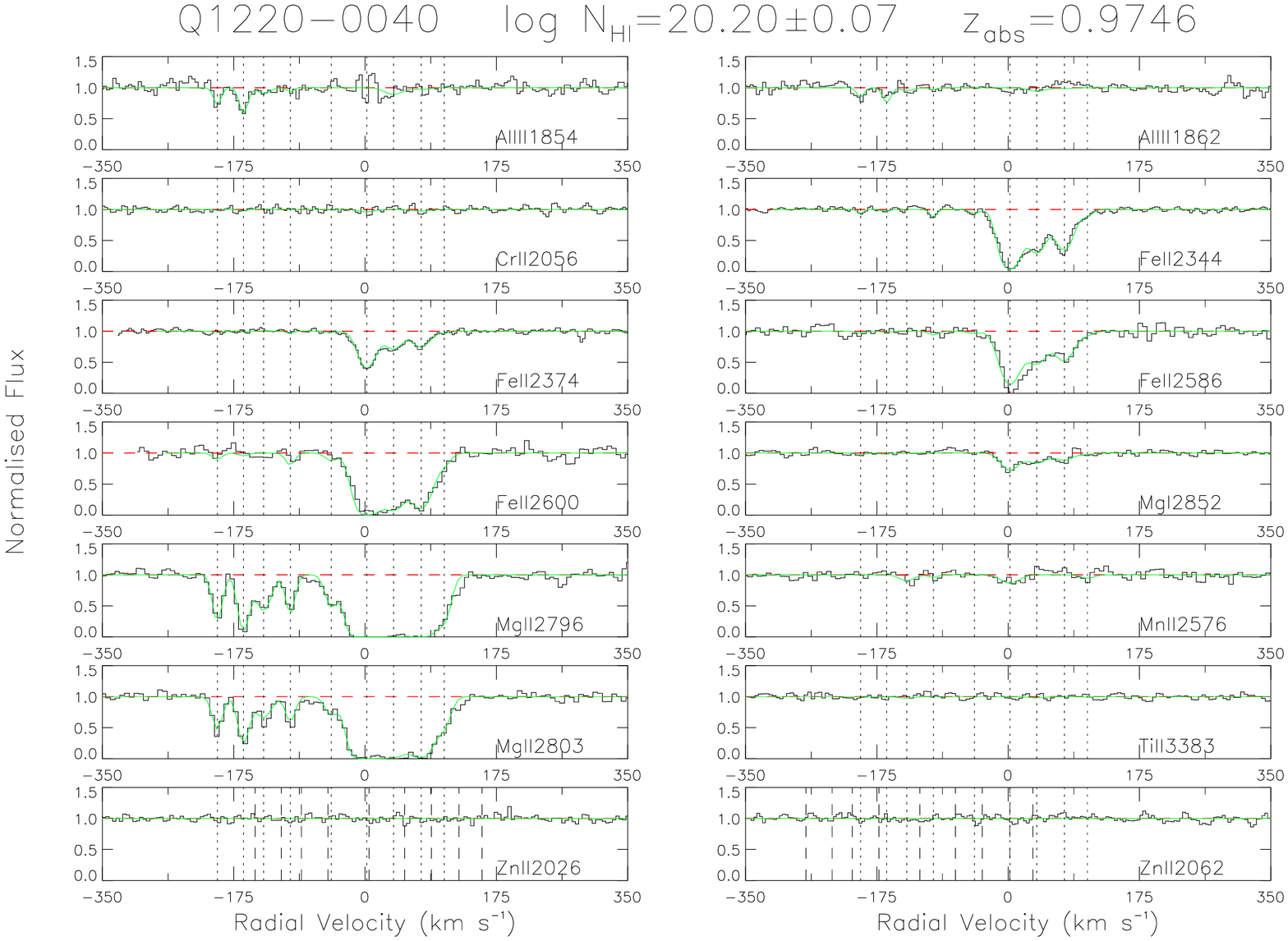} \\ [0.5cm] \includegraphics[angle=90, width=5.5in]{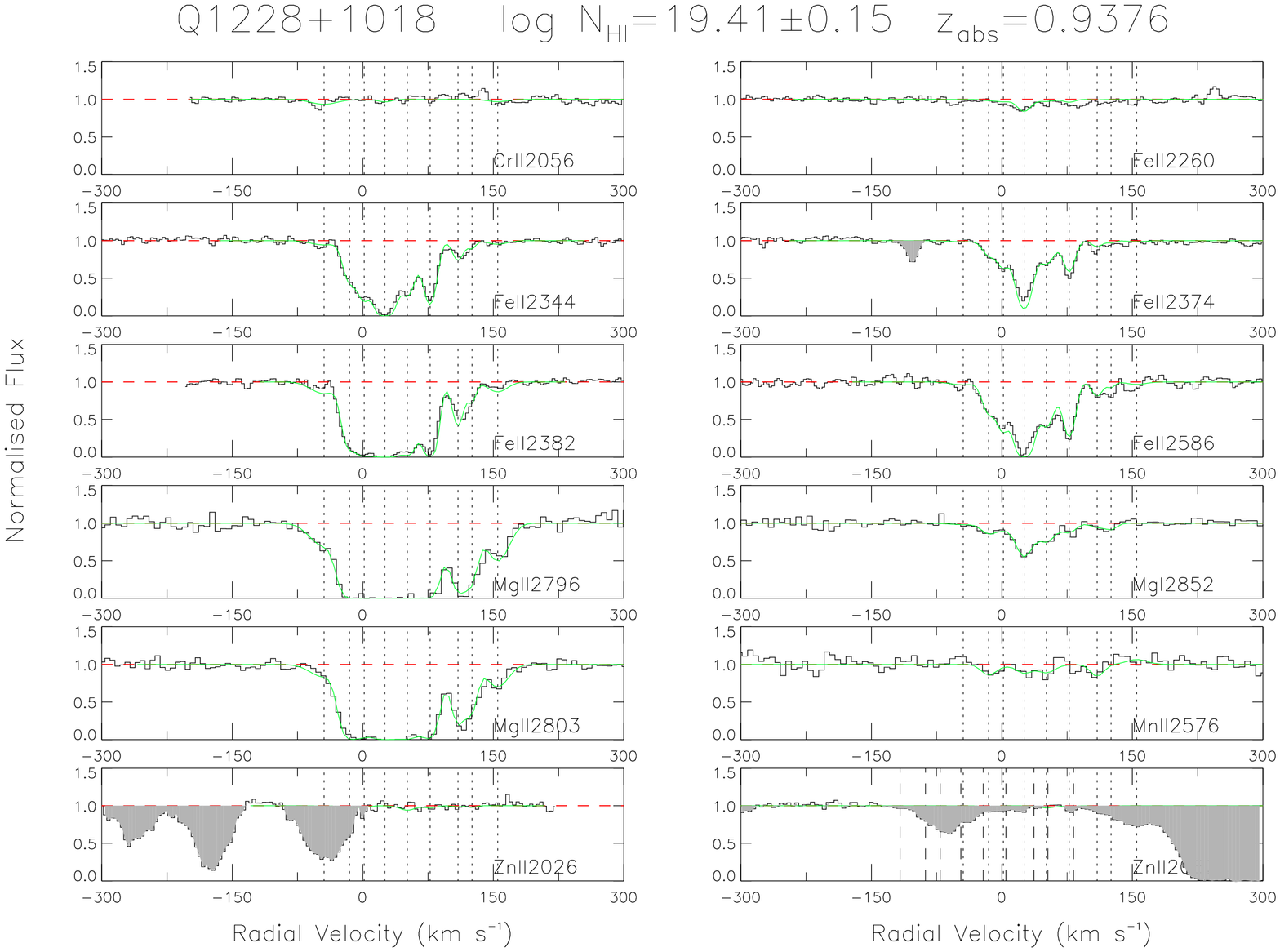} \\ [0.0cm]

  \end{array}$
  \end{center}
  \small{\textbf{ Figures 5 and 6:} Same as Figure 1, but for the \za=0.9746 system in Q1220-0040 (top), and the \za=0.9376 system in Q1228+1018 (bottom). The areas shaded in 
  grey in the Zn II \lala 2026, 2062 panels for  the \za=0.9376 system in Q1228+1018 represent absorption from lines inside the Lyman-$\alpha$ forest, and are not due to 
  absorption from Zn II. There is a small, unidentified feature near the Fe II $\lambda$ 2374 line, which has also been shaded in grey.}
  \end{minipage}
  \end{figure*}

\begin{figure*}
  \begin{minipage}{160mm}
  \begin{center}
 $\begin{array}{c@{\hspace{0.0in}}c}
  \multicolumn{1}{l}{\mbox{\bf }} &
  	\multicolumn{1}{l}{\mbox{\bf }} \\ [0.0cm]
  		\includegraphics[angle=90, width=5.5in]{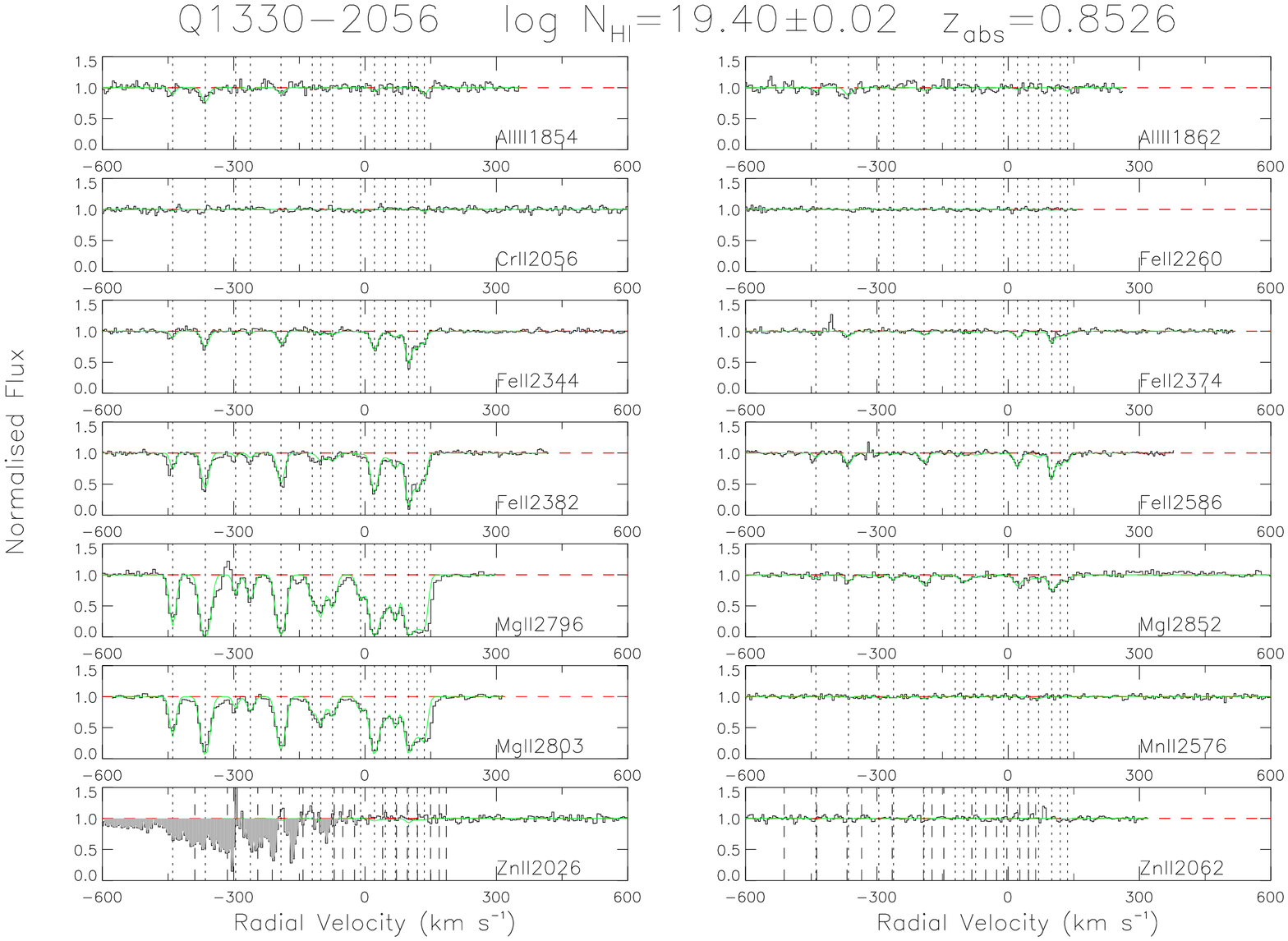} \\ [0.5cm] \includegraphics[angle=90, width=5.5in]{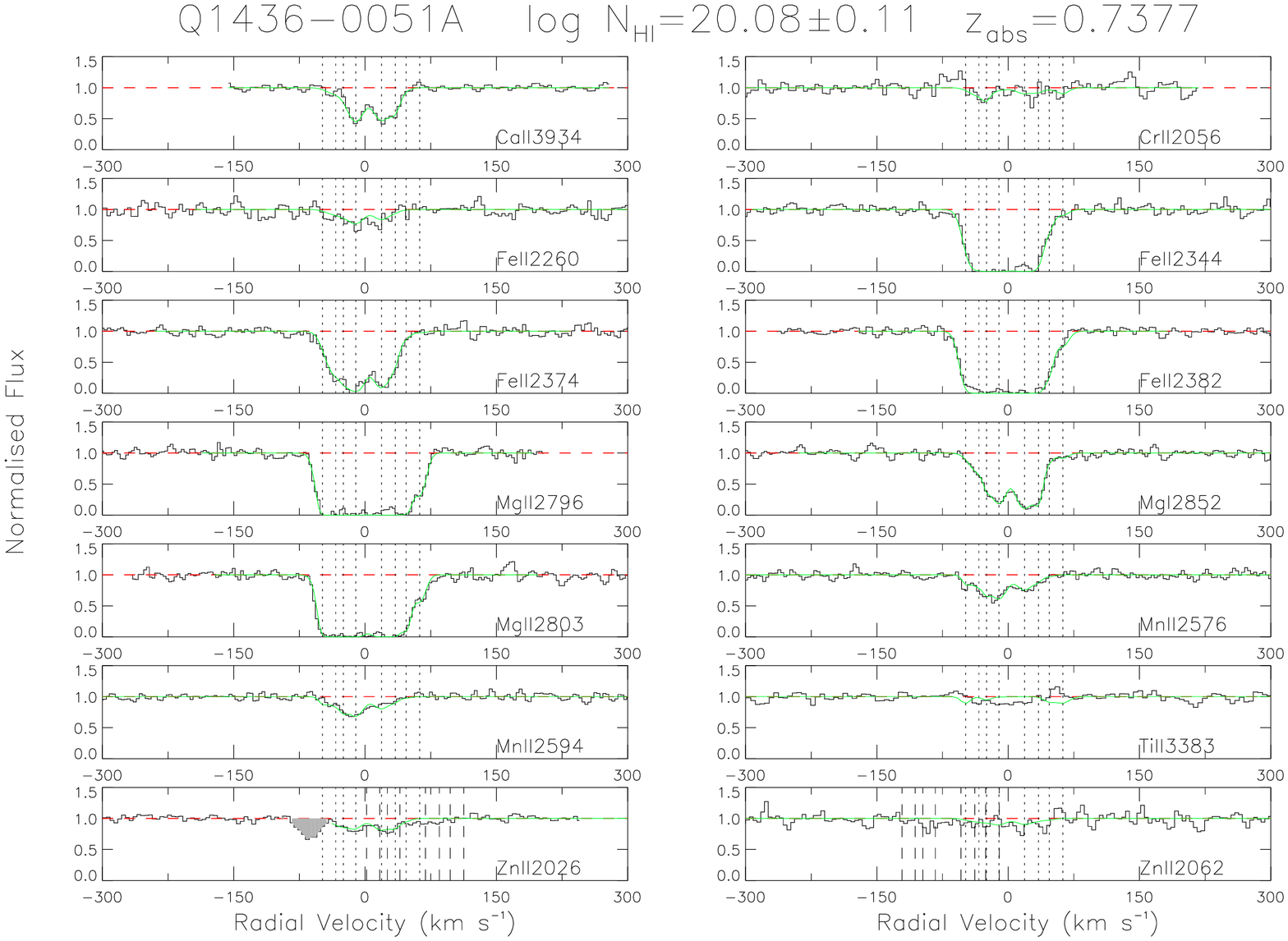} \\ [0.0cm]

  \end{array}$
  \end{center}
  \small{\textbf{ Figures 7 and 8:} Same as Figure 1, but for the \za=0.8526 system in Q1330-2056 (top), and the \za=0.7377 system in Q1436-0051 (system A, bottom). 
 For the \za=0.8526 system in Q1330-2056, in the region around the expected position of Zn II $\lambda$ 2026, there is a bad column in the blue CCD that 
  caused the feature seen in the figure. The part of the profile that is affected by the bad column is shaded in grey to highlight this defect. 
  For the \za=0.7377 system in Q1436-0051, we have highlighted the contribution of the C IV $\lambda$ 1550 line at \za=1.2699 to the Zn II 2026 profile in grey.
   See $\S$ 4.7 for a discussion of this blend.  }
  \end{minipage}
  \end{figure*}

\begin{figure*}
  \begin{minipage}{160mm}
  \begin{center}
 $\begin{array}{c@{\hspace{0.0in}}c}
  \multicolumn{1}{l}{\mbox{\bf }} &
  	\multicolumn{1}{l}{\mbox{\bf }} \\ [0.0cm]
  		\includegraphics[angle=90, width=5.5in]{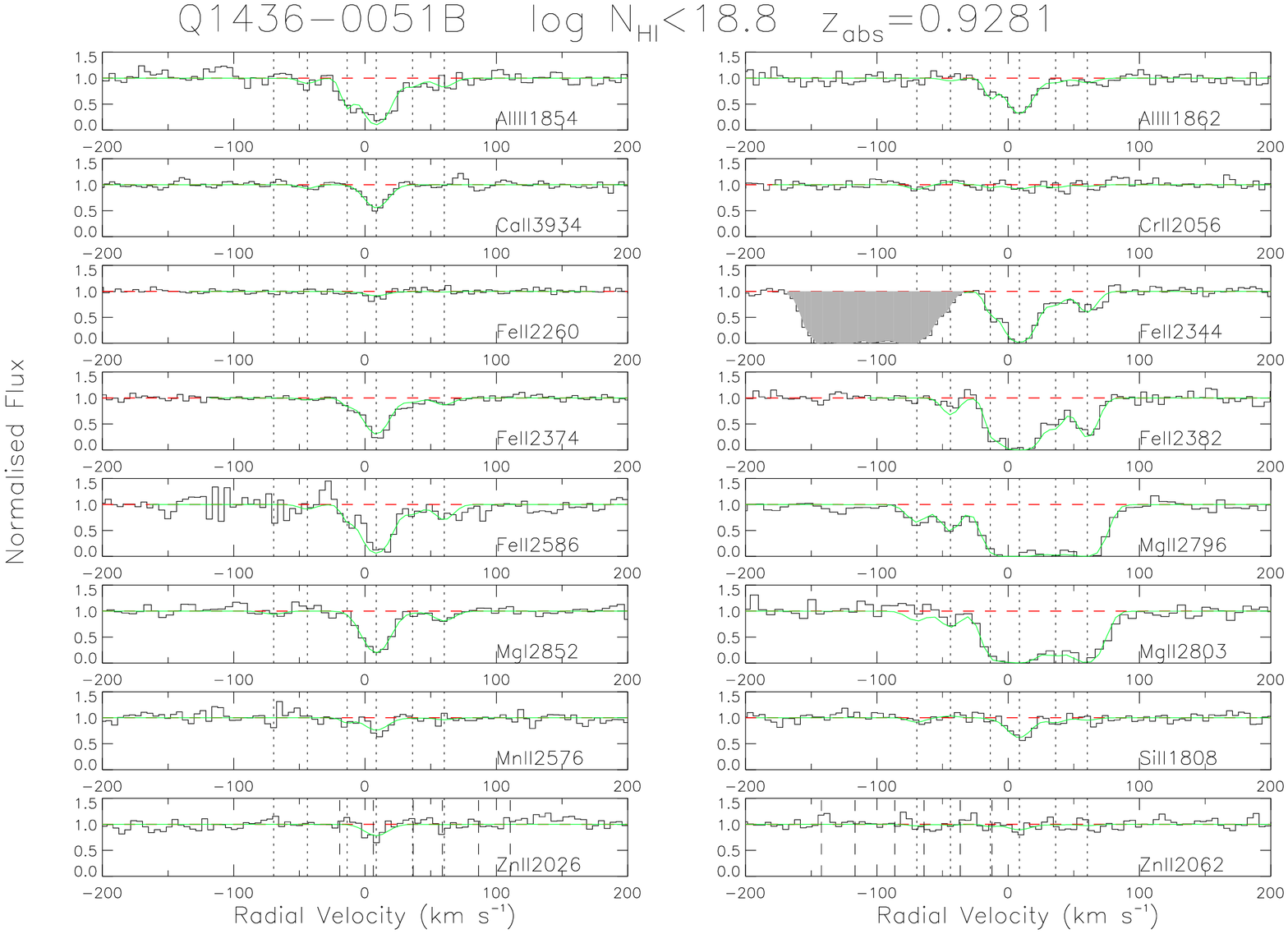} \\ [0.5cm] \includegraphics[angle=90, width=5.5in]{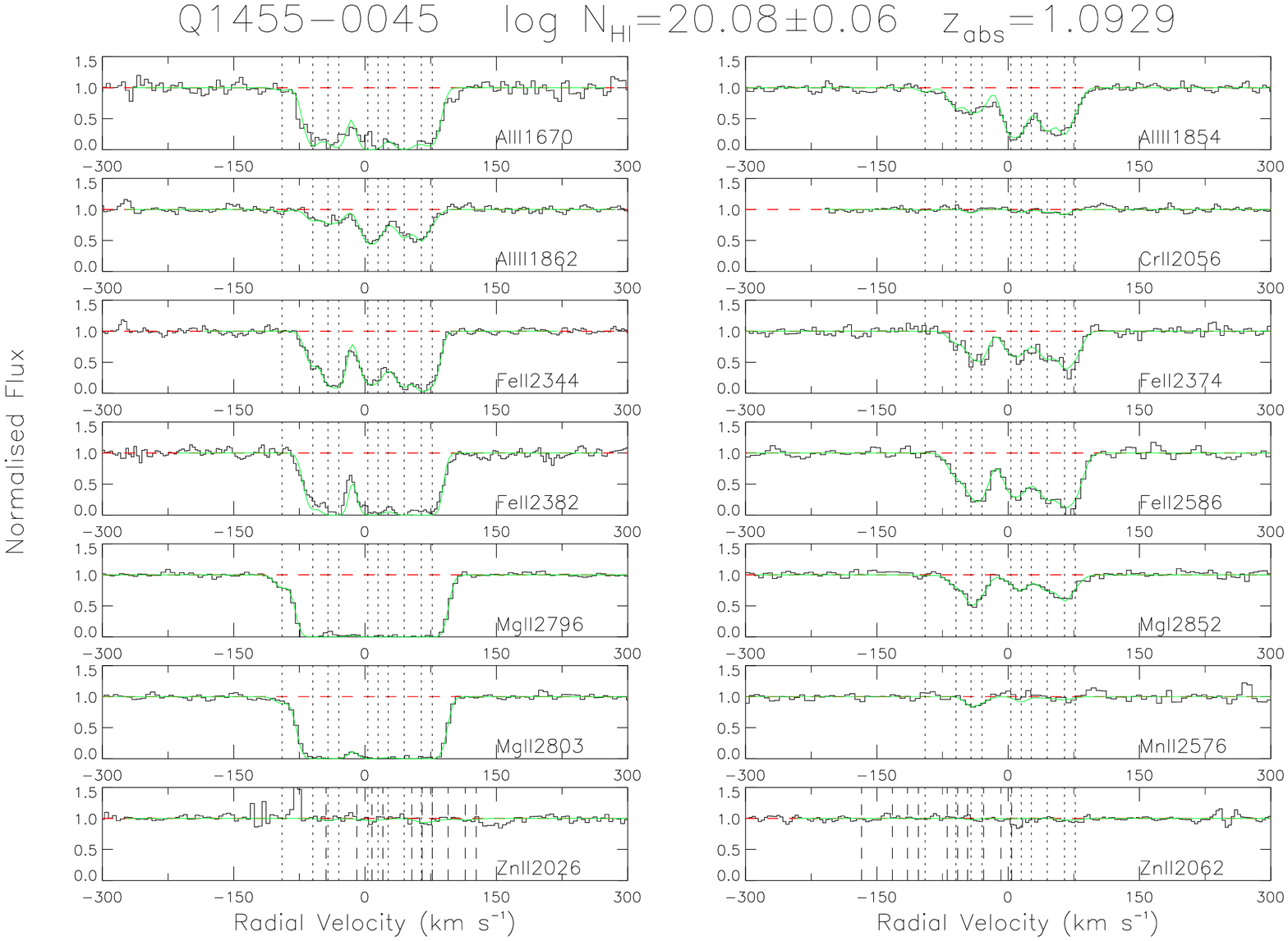} \\ [0.0cm]

  \end{array}$
  \end{center}
  \small{\textbf{Figures 9 and 10:} Same as Figure 1, but for the \za=0.9281 system in Q1436-0051 (system B, top), and the \za=1.0929 system in Q1455-0045 (bottom).  }
  \end{minipage}
  \end{figure*}

\subsection{Q1436-0051 (z$_{em}$=1.275)}
\textbf{(System A:\za=0.7377):}
This is a sub-DLA system with log \nhI=20.08$\pm$0.11 \citep{Rao06}. 
We detect absorption lines from Mg I, Mg II, Ca II, Mn II, Fe II, and Zn II in system A. A total of 8 components were used to fit the absorption profiles. There is also a C IV system
at \za=1.2699, with the C IV $\lambda$ 1550 line close to the expected position of the Zn II $\lambda$ 2026 line in the sub-DLA at \za=0.7377, 
as can be seen in the velocity plots shown in Figure 8. The C IV 
\lala 1548, 1550 lines were fit simultaneously, and later held fixed while fitting the Zn II \lala 2026, 2062 lines simultaneously. 
The C IV system shows a simple structure with only 2 components needed to fit the profile, with no absorption in the region where the majority of the Zn II $\lambda$ 2026 absorption is.
 In Figure 8, the C IV contribution to this line is highlighted in grey.
The Mg I contribution to the Zn II $\lambda$ 2026 line, based on the fit to the Mg I 2852 line, was also included and held fixed. 
This was the only case of an appreciable contribution of Mg I to the blended Zn II + Mg I $\lambda$ 2026 line that was seen.
This system has a high metallicity with
[Zn/H]$=-$0.05$\pm$0.11, and [Fe/H]$=-$0.61$\pm$0.11. 
We also detect strong Mn II \lala 2576, 2594, 2606 lines with [Mn/H]$=-$0.57$\pm$0.11 and  [Mn/Fe]=+0.04$\pm$0.03. 
The Ca II $\lambda$ 3934 line was detected with W$_{0}$=396$\pm$15 m\ang, but the Ca II $\lambda$ 3969 line was blended with telluric features. 
The Si II $\lambda$ 1808 line was not observable for this sub-DLA, so no information on $alpha$-enhancement is available for this system. 
We show the results of the profile fitting analysis in table 11, and give the rest frame equivalent widths for this system in table 2.

\textbf{(System B:\za=0.9281):} This is a Lyman-limit system with log \nhI$<$18.8 \citep{Rao06}. 
Several lines are detected in this system, such as Mg I, Mg II, Al III, Si II, Ca II, Mn II, and Fe II. 
A total of 6 components were used in the Voigt profile fits for this system, although Si II, Ca II, and Zn II were only detected in the $v\sim$9 \kms component.
This system has a large amount of Al III and likely has a high degree of ionisation, although the Al II $\lambda$ 1670 line was not in the wavelengths covered. 
The Ca II $\lambda$ 3969 line was again blended with telluric features. The Fe II $\lambda$ 2344 
line is partly contaminated by the Fe II $\lambda$ 2600 line from system A. 

This system appears to have a super-solar abundance, 
with [Zn/H]$>$+0.86 and based on Fe II, the metallicity is [Fe/H]$>-$0.07. As was mentioned in $\S$ 4.2, the [Fe/H] may be lower than this due to ionisation corrections.
This system also shows a super-solar abundance in both Mn with
[Mn/H]$>$+0.15 and Si with [Si/H]$>$+0.68. If the Mg I contribution to the blended Zn II + Mg I $\lambda$ 2026 line is substantially more than what we have estimated based on the Mg I $\lambda$ 2852 
line, then the Zn II metallicity could be overestimated. However, the component structure of Mg I $\lambda$ 2852 does not suggest that an Mg I $\lambda$ 2026
blend could affect the detected dominant Zn II $\lambda$ 2026 component.
 With such a low H I column density, it is possible that a substantial ionisation correction factor is
needed for this system. The Al III/Al II ratio which can be used to estimate ionisation, could not be determined for this system as the Al II $\lambda$ 1670 was not covered. 
Even with $-$1.0 dex ionisation corrections lowering the abundances, this system would still be well above the average metallicity of DLAs.
 Figure 9 gives velocity plots of several lines of interest. We give the results of the profile fitting analysis in table 12, along with the rest frame equivalent widths for the lines in this 
 system in table 2. 

\begin{table*}
\begin{minipage}{135mm}
\begin{center}
\caption{ Same as table 3, but for the \za=0.7377 system in  Q1436-0051 with \nhI=20.08$\pm$0.11.}
\begin{tabular}{cccccccc}
\hline
\hline
v	&b$_{eff}$	&	Mg I 			&	Mg II			&	Ca II			&	Mn II			&	Fe II			&	Zn II			\\
\hline
$-$49	&	3.6	&	-			&	$>$1.00E13		&	-			&	(4.40$\pm$1.07)E11	&	-			&	-			\\
$-$34	&	13.2	&	(6.65$\pm$0.76)E11	&	$>$8.68E13		&	(4.18$\pm$1.37)E11	&	(2.18$\pm$0.21)E12	&	(1.56$\pm$0.06)E14	&	-			\\
$-$25	&	3.5	&	(2.67$\pm$0.81)E11	&	-			&	-			&	(6.66$\pm$1.53)E11	&	(2.09$\pm$0.12)E13	&	-			\\
$-$11	&	11.1	&	$>$2.51E12		&	$>$5.74E13		&	(2.69$\pm$0.17)E12	&	(4.11$\pm$0.19)E12	&	(3.82$\pm$0.33)E14	&	(2.17$\pm$0.49)E12	\\
19	&	11.5	&	$>$3.48E12		&	$>$7.13E13		&	(2.61$\pm$1.68)E12	&	(2.47$\pm$0.16)E12	&	(2.70$\pm$0.16)E14	&	(1.74$\pm$0.17)E12	\\
35	&	5.6	&	$>$1.31E12		&	$>$2.26E13		&	(9.22$\pm$1.25)E12	&	(4.29$\pm$1.12)E11	&	(3.00$\pm$0.34)E13	&	(7.16$\pm$1.36)E11	\\
47	&	5.7	&	(1.15$\pm$0.36)E11	&	$>$1.23E13		&	-			&	-			&	(6.94$\pm$0.71)E12	&	-			\\
62	&	8.0	&	-			&	(4.03$\pm$0.21)E12	&	-			&	-			&	(2.24$\pm$0.52)E12	&	-			\\
\hline
\end{tabular}
\end{center}
\end{minipage}
\end{table*}

\begin{table*}
\begin{minipage}{165mm}
\begin{center}
\caption{ Same as table 3, but for the \za=0.9281 system in Q1436-0051 with \nhI$<$18.8.}
\begin{tabular}{cccccccccc}
\hline
\hline
v	&b$_{eff}$	&	Mg I 			&	Mg II			&	Al III				&	Si II			&	Ca II			&	Mn II			&	Fe II			& Zn II	\\
\hline
$-$70	&	6.8	&	-			&	(1.70$\pm$0.37)E12	&	-				&	-			&	-			&	-			&	-			&	-	\\
$-$44	&	6.6	&	-			&	(3.00$\pm$0.51)E12	&	-				&	-			&	-			&	-			&	-			&	-	\\
$-$14	&	2.7	&	-			&	$>$1.09E15		&	(4.62$\pm$0.93)E12		&	-			&	-			&	(4.64$\pm$1.82)E11	&	(8.56$\pm$1.00)E12	&	-	\\
9	&	10.7	&	(2.79$\pm$0.28)E12	&	$>$9.99E13		&	(1.83$\pm$0.14)E13		&	(1.04$\pm$0.13)E15	&	(1.89$\pm$0.16)E12	&	(2.36$\pm$0.27)E12	&	(1.30$\pm$0.05)E14	&	(1.97$\pm$0.27)E12	\\
36	&	7.3	&	-			&	$>$1.61E13		&	(1.03$\pm$0.33)E12		&	-			&	-			&	-			&	(6.91$\pm$0.70)E12	&	-	\\
60	&	8.0	&	(2.86$\pm$0.75)E11	&	$>$1.15E14		&	(1.17$\pm$0.33)E12		&	-			&	-			&	-			&	(1.26$\pm$0.08)E13	&	-	\\
\hline
\end{tabular}
\end{center}
\end{minipage}
\end{table*}

\begin{minipage}{115mm}
\end{minipage}

\begin{minipage}{115mm}
\end{minipage}

\subsection{Q1455-0045 (z$_{em}$=1.378)}
\textbf{(System A:\za=1.0929):} There is a sub-DLA system with log \nhI=20.08$\pm$0.06 in the spectrum of this QSO \citep{Rao06}. 
We detected strong absorption lines from Mg I, Mg II, Al II, Al II, Si II, and Fe II
in this system. A total of 10 components were used in the Voigt profile fits. No Zn II \lala 2026, 2062 lines were detected at S/N$\sim$27 in the region. We could constrain the 
metallicity based on Zn as [Zn/H]$<-$0.80. The abundances based on the Fe II and Si II lines are [Fe/H]$=-$0.98$\pm$0.06 and [Si/H]$=-$0.98$\pm$0.12. 
The Mn II $\lambda$ 2576 was also detected with [Mn/H]=$-$1.43 and [Mn/Fe]$=-$0.51$\pm$0.15
We did detect both the Al II and Al III lines, with the ratio of column densities Al III/Al II $<-$0.18. 
Weak features of Cr II $\lambda$ 2056 were also detected with W$_{0}$=21$\pm$7 m\ang. Only one component at $\sim$64 \kms was above a 3$\sigma$ significance level 
with [Cr/Fe]$=-$0.35$\pm$0.13, although another component at $\sim$ 44 \kms was at the $\sim$2$\sigma$ level. 
The Cr II column density was found to be higher via the AOD method, with [Cr/Fe]=0.0. Figure 10 shows velocity plots of several lines of interest. We show the results of the profile fitting
 analysis in table 13, and the rest frame equivalent widths for this system in table 2.

\begin{table*}
\begin{minipage}{165mm}
\begin{center}
\caption{ Same as table 3, but for the \za=1.0929 system in the spectrum of Q1455-0045 with \nhI=20.08$\pm$0.06.}
\begin{tabular}{cccccccccc}
\hline
\hline
v	&b$_{eff}$	&	Mg I			&	Mg II			&	Al II		&	Al III			&	Si II			&	Cr II		&	Mn II			&	Fe II	\\																																																																																																																																																																																																																																														
\hline
$-$95	&	8.8	&	-		&	(1.00$\pm$0.09)E12		&	-		&	(4.92$\pm$1.51)E11	&	-			&	-		&	-			&	-	\\																																																																																																																																																																																																																																														
$-$60	&	9.8	&	(3.01$\pm$0.50)E11	&	$>$2.02E14		&	$>$7.21E12	&	(2.87$\pm$0.22)E12	&	-			&	-		&	-			&	(2.15$\pm$0.09)E13	\\																																																																																																																																																																																																																																														
$-$42	&	7.5	&	(7.42$\pm$0.78)E11	&	$>$8.96E12		&	$>$3.09E12	&	(2.46$\pm$0.23)E12	&	-			&	-		&	(1.23$\pm$0.40)E12	&	(3.64$\pm$0.20)E13	\\																																																																																																																																																																																																																																														
$-$30	&	8.2	&	(4.12$\pm$0.59)E11	&	$>$1.33E14		&	$>$7.51E12	&	(2.37$\pm$0.22)E12	&	-			&	-		&	-			&	(5.02$\pm$0.22)E13	\\																																																																																																																																																																																																																																														
3	&	10.3	&	(1.81$\pm$0.55)E11	&	$>$1.10E14		&	$>$1.62E13	&	(1.06$\pm$0.05)E13	&	-			&	-		&	-			&	(4.73$\pm$0.19)E13	\\																																																																																																																																																																																																																																														
15	&	7.1	&	(2.95$\pm$0.65)E11	&	$>$1.05E13		&	$>$5.51E12	&	(4.69$\pm$0.50)E12	&	-			&	-		&	-			&	(2.24$\pm$0.20)E13	\\																																																																																																																																																																																																																																														
26	&	8.1	&	-			&	$>$2.20E13		&	$>$1.80e12	&	(2.03$\pm$0.30)E12	&	-			&	-		&	-			&	(1.51$\pm$0.13)E13	\\																																																																																																																																																																																																																																														
45	&	11.1	&	(4.89$\pm$0.59)E11	&	$>$7.02E13		&	$>$1.67E12	&	(9.50$\pm$0.44)E12	&	-			&	-		&	-			&	(6.78$\pm$0.26)E13	\\																																																																																																																																																																																																																																														
64	&	7.9	&	(6.30$\pm$0.71)E11	&	$>$5.66E13		&	$>$3.05E12	&	(7.52$\pm$0.46)E12	&	-			& (2.51$\pm$0.71)E12 	&	-			&	(6.66$\pm$0.38)E13	\\																																																																																																																																																																																																																																														
77	&	8.8	&	(2.17$\pm$0.53)E11	&	$>$2.18E14		&	$>$6.12E12	&	(2.56$\pm$0.25)E12	&	(4.40$\pm$0.10)E14	&	-		&	-			&	(4.65$\pm$0.21)E13	\\																																																																																																																																																																																																																																														
\hline
\end{tabular}
\end{center}
\end{minipage}
\end{table*}

\begin{minipage}{115mm}
\end{minipage}

\section{Discussion and Conclusions}

The abundances for the observed systems are given in table 14.
We have used the total column densities (i.e. the sum of the column densities in the individual components of a system that were determined via profile fitting method)
along with the total \nhI as given in table 1, to determine the abundances of these systems. It would be interesting to determine the abundances in the individual components, as
they may possibly vary from component to component. This however is impossible at the present time due to the lack of high resolution UV spectrographs that would be necessary 
to determine \nhI in individual components. We have ignored any ionisation corrections while determining these abundances, and have assumed the first ions to be the dominant 
ionisation species of the elements for which these abundances have been determined, namely Zn, Fe, Mn, Cr, and Si. We discuss the issue of ionisation further below.
Solar systems abundances from \citet{Lodd03} are also given in table 14. The abundances for Ca II are likely only lower limits, as the ionisation potential of 11.868 eV for Ca$^{+}$ is lower than 
ionisation potential of H$^{0}$ (13.6 eV). 

Relative abundances of various elements are also given in table 14, with the column densities determined from the profile fitting analysis. We give the metallicities of these systems, along
with the [Zn/Fe] ratio, which is often used as an indicator of dust depletion. We also provide the ratios of [Si/Fe], [Ca/Fe], [Cr/Fe], and [Mn/Fe]. 
Finally, we provide ratios of the column densities of the adjacent ions Al III/Al II and Mg II/Mg I as well as Mg II/Al III and Fe II/Al III,
any of which may provide information about the ionisation in these
systems. By matching the observed ratios of the ions mentioned above to those generated through grids of photoionisation models,
one can constrain the levels of ionisation, and any ionisation correction factors for the observed abundances.

Based on grids of Cloudy models, is has previously been shown that there appears to be little need for ionisation corrections in most sub-DLA systems \citep{Des03, Mei07}. 
We have therefore not applied any ionisation corrections to to the abundances in table 14. We do note however, that the \za=0.8301 system  with \nhI=18.95$\pm$0.18 in Q1054-0020 and the
\za=0.9281 system with \nhI$<$18.8 in Q1436-0051 both show possible signs of high levels of ionization, with 
Fe II/Al III=0.74 and Fe II/Al III=0.80 respectively. 
Substantial ionisation corrections could be necessary. Cloudy models for the \za=0.9281 system in Q1436-0051 indicate a positive ionization correction factor for Zn
in this system of $\sim$+0.2 to +0.3 dex for $-6 <$log U$<-3$ (indicating that [Zn II/H I] underestimates the true abundance by this amount).
 We caution that these results may be inaccurate due to uncertainties in the atomic data for Zn.  The ionisation correction factors
for Si, Mn, and Fe are all negative over this range of ionisation parameters with values of $\sim-$0.3 to $-$0.5 dex for Mn and Fe, and values ranging from $\sim$0.0 to $\sim-$1.25 for Si. 
Based on these models, it appears that [Fe/Zn] is suppressed as one goes to lower and lower \nhI values, 
the true depletion is actually much higher than it appears based on Zn II and Fe II. Similar values were derived for the \za=0.8301 system  with \nhI=18.95$\pm$0.18 in Q1054-0020 based on Cloudy models. 
As no adjacent ions that could constrain the ionisation parameter are available for these systems, we leave the abundances unaltered in table 14. 
We prefer to leave the corrections to a later time when more constraints are available (i.e. when we have UV spectra with coverage of Fe III, S III, Si IV, etc and better 
atomic data become available).

Variations by component of the relative abundances of elements can also provide information about the local star formation history and possibly ionization. \citet{Pro02} 
noticed a remarkably small ($\sim$0.3 dex) component-by-component dispersion in [Zn/Fe] over a wide range of metallicities. 
As we only have Zn detections for two of the systems here, we instead look at [Mn/Fe], which is likely less affected by dust
depletion (Mn and Fe have similar condensation temperature) and more affected by nucleosynthetic enrichment. Unfortunately, other ions such as Mg II and Al II are too saturated to determine accurate
column densities in most of the individual components. The \za=1.4244 system in Q1037+0028 and the \za=0.8301 system in Q1054-0020 show 
similarly low dispersions of [Mn/Fe], but interestingly the high metallicity \za=0.7377 and \za=0.9281 systems of Q1436-0051 show higher spreads of [Mn/Fe] of $\sim$0.6 and $\sim$0.5 dex 
respectively. The metal rich sub-DLA observed in \citet{Per06a} also showed a large dispersion in [Mn/Fe]. The \za=0.7377 system in Q1436-0051 also shows a 0.6 dex range of [Zn/Fe] in the components.  
One possible explanation for these larger dispersions could be enrichment by recent supernovae explosions in which the gas has not had time to sufficiently mix.

Zn II was detected in only 2 of the systems, and upper limits on the Zn abundance were placed on the rest of the systems. The Cr II $\lambda$ 2056 line was detected only in the 
\za=1.0929 system in Q1455-0045. Due to the weakness of the Lyman-$\alpha$  line in the \za=0.9281 system in Q1436-0051, only an upper limit could be placed on the neutral hydrogen column density, 
and therefore lower limits on the abundances in this system. This system shows an unusual abundance pattern, with high metallicity ([Zn/H]$>$+0.86, [Fe/H]$>-$0.07, and [Si/H]$>$+0.68) 
and high depletion ([Zn/Fe]=+0.94). This system also showed an overabundance of Mn, with [Mn/Fe]=+0.22. 
As Mn is an odd element, and Fe is even, [Mn/Fe] might be below solar in material whose source is dominated by the products of Type II supernovae, but likely not higher if 
differential dust depletion effects are negligible. In the other case of a Zn detection, the system at \za=0.7377 in Q1436-0051, Mn is also slightly more abundant relative to Fe, with 
[Mn/Fe]=+0.04. \citet{Led02} found a large dispersion in [Mn/Fe] based on a sample of DLA absorbers and concluded that this large range of values not seen in other abundance ratios 
such as [Cr/Fe] or [Si/Fe] must be due to a combination of both depletion and nucleosynthetic effects.
 
A similar over-abundance of Mn relative to Fe was also seen by \citet{Per06a} in another sub-DLA at \za=0.716 with [Zn/H]=+0.61 and [Mn/Fe]=+0.14. This could perhaps be due to
heavy depletion of Fe compared to Mn onto dust grains, unmixed gas that has recently been expelled through stellar winds and supernovae, or an intrinsically non-solar abundance pattern.
Cloudy models for this system indicate that the corrections for [Mn/Fe] due to ionisation are small ($\la$ 0.10 dex for $-6<$log U$<-1$). 
In the other 3 cases where we have detections of Mn and Fe, but no Zn, Mn is slightly under-abundant relative to Fe, with $<$[Mn/Fe]$>$=$-$0.31. The systems in our sample have an average 
Fe abundance of $<$[Fe/H]$>$=$-$0.68, which is higher than for typical DLA systems ($<$[Fe/H]$>_{\rm DLA}\sim-$1.5). 
This may be due to less dust depletion or truly higher abundances in these systems, or to undue influence of ionisation effects in the low \nhI systems. 

To determine an average [Zn/H] abundance of the absorbers in this sample, the spectra were shifted to rest wavelengths and averaged. 
Any obvious non-associated features in the region around the Zn II $\lambda$ 2026 line in the individual spectra were replaced by random noise with S/N similar to the uncontaminated regions.
We show the average spectrum around Zn II $\lambda$ 2026 \ang in Figure 11. A feature was seen in the region with \w=8$\pm$2 m\ang and log N$_{\rm Zn \ II}$=11.63. The feature 
at $\sim$+70 \kms may be due to absorption from Mg I $\lambda$ 2026.477, and was not included when determining \w. 
The systems in this sample have an average H I column density of log $<$N$_{\rm H \ I}$$>$=19.80, giving this sample an average Zn abundance of $<$[Zn/H]$>=-$0.80 based on this composite spectrum.
To increase the S/N, we also binned the averaged spectrum to a dispersion of 0.5 \ang/pixel ( R$\sim$4000 ). 
After binning, a feature was detected at a $\ga$2$\sigma$ significance level with \w=11m\ang. If the feature is assumed to be a true detection, 
the Zn II column density is log N$_{\rm Zn \ II}$=11.80, with a corresponding mean abundance of $<$[Zn/H]$>$=$-$0.63 for these absorbers at 0.7 $\la$ \za$ \la$ 1.5 based on the binned spectrum.
In comparison, DLA absorbers have a mean Zn abundance of $<$[Zn/H]$>$$\sim-$1.0 in this redshift range \citep{Kul05, Kul07}. 

The \nhI-weighted mean metallicity given by: 
\begin{equation}\rm{[\langle X/H \rangle]=log \langle (X/H) \rangle - log(X/H)_{\sun}}\end{equation}
where
\begin{equation}\rm{ \langle(X/H)\rangle=\frac{\sum_{i=1}^nN(X)_{i}}{\sum_{i=1}^nN(H)_{i}}}\end{equation}
has often been used as a quantitative way of estimating the amount of metal enrichment in the Universe in a given epoch \citep{Kul05, Kul07}
For the absorption systems in this sample, the  \nhI-weighted mean metallicity based on Fe is [$<$Fe/H$>$]=$-$0.76. Using Zn which is less affected by dust depletion, we determine the 
\nhI-weighted mean metallicity for these systems at 0.7 $\la$ \za$ \la$ 1.5. to be [$<$Zn/H$>$]=$-$0.40 assuming the upper limits to be detections, 
and [$<$Zn/H$>$]=$-$0.70 assuming the limits to be zero.

\setcounter{figure}{10}
\begin{figure}
\includegraphics[height=\linewidth,  angle=90]{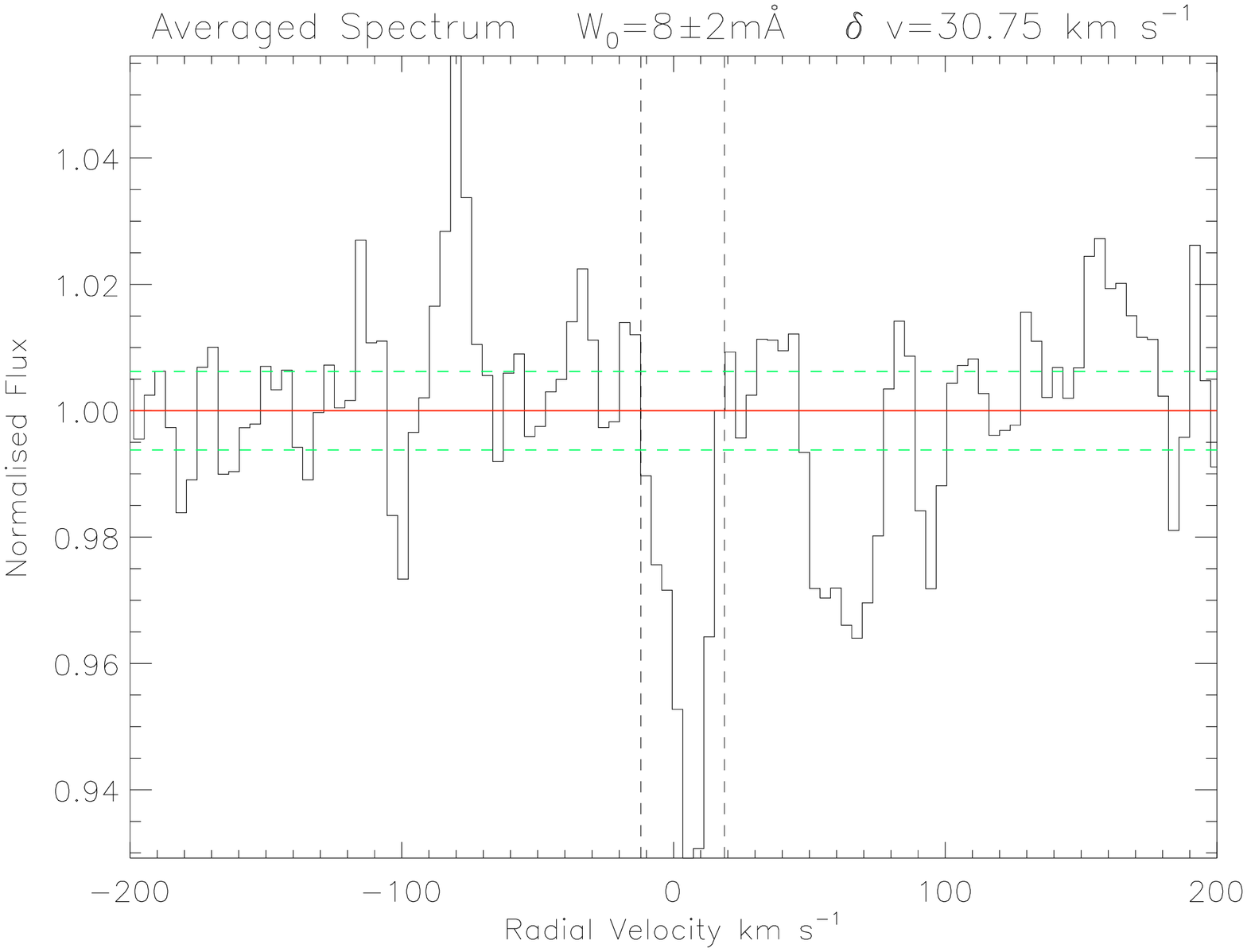}
\caption{The  unbinned average spectrum of the 10 absorbers in this sample near Zn II $\lambda$ 2026 \ang. The horizontal red line indicates the continuum level, and the horizontal dashed green lines
show the best fit continuum level increased or decreased by $\pm$40$\%$ of the RMS variance in the noise. 
The vertical dashed lines show the integration limits. The feature seen was detected at $\ga4\sigma$ with \w=8$\pm$2 m\ang. }
\end{figure}

In this paper we have presented medium-resolution spectra of 10 sub-DLAs at \za$\la$1.5. Although to date the DLA systems have been the preferred tracer of metallicity 
at high redshift, every absorber with solar or higher metallicity that has been detected has been a sub-DLA \citep{Pet00, Kh04, Per06a, Pro06, Mei07}. With this latest sample of
sub-DLAs, we have found two more systems with high metallicity, [Zn/H]$=-$0.05 and [Zn/H]$>$+0.86
Two other systems in this sample that did not have a Zn II detection, did nonetheless have [Fe/H]$>-0.30$. 

These observations suggest that the largely ignored sub-DLAs may provide
important clues into the chemical evolution of the Universe. Clearly, not all sub-DLAs have solar or higher metallicity, nor would one expect them to if they trace 
typical galaxies which have a range of metallicities. 
Sub-DLAs do appear, however, to have a higher mean metallicity  and show faster evolution than do DLAs \citep{Kul07}. 
We continue to find systems with low \nhI and high metallicities, but higher S/N is needed to determine if the systems with upper limits of [Zn/H]$\sim$-0.3
 are indeed at that level, or are much less, comparable to DLA abundances.
With our sample sizes of sub-DLAs now increasing, we can begin to address some of the questions relating to the chemical evolution of metals. Future UV spectra would also allow measurements
of key lines of S III and Fe III which could yield important information about ionisation. In forthcoming work, and with an extended sample, we will
examine differences in the properties of sub-DLAs and DLAs, including abundances and kinematical structure, as well as the redshift evolution of these properties in sub-DLAs.

\begin{sidewaystable*}
\begin{minipage}{220mm}
\textbf{Table 14.} Relative abundances and abundance ratios for the observed sub-DLAs. The first line below the heading gives the solar reference abundance for that column. 
\begin{tabular}{lccccccccccccc}
\hline
\hline
QSO		&	\za		&	Log \nhI	&	[Zn/H]	 		&	[Fe/H]			&	[Zn/Fe]			&	[Si/Fe]		&	[Ca/Fe]			&	[Cr/Fe]		&	[Mn/Fe]			&Al III/Al II$^{a}$ 	&	Mg II/Mg I$^{a}$&	Mg II/Al III$^{a}$	&	Fe II/Al III$^{a}$\\
 \hline	 																											
[X/Y]$_{\sun}$	&			&			&	$-$7.37			&	$-$4.53			&	$-$2.85			&	+0.07		&	$-$1.13			&	$-$1.82		&	$-$1.97			&			&			&				&			\\
\hline																											
 Q1037+0028	& 	1.4244	 	& 	20.04$\pm$0.12 	&	$<-0.63$		& 	$-$0.67$\pm$0.12 	&	$<+0.04$		& 	0.18$\pm$0.04	&	 	-	 	&	$<-0.76$	&	 $-$0.34$\pm$0.05 	&	$<-0.96$	&	 -	 	&	$>$2.29		 	&	1.65$\pm$0.03	\\
 Q1054-0020A	& 	0.8301	 	& 	18.95$\pm$0.18 	&	$<+0.18$		& 	$-$0.09$\pm$0.18 	&	$<+0.28$		&	 	-	&	 	-	 	&	$<-0.04$	&	 $-$0.09$\pm$0.07 	&	 	- 	&	$>$2.28	 	&	$>$1.24		 	&	 0.74$\pm$0.06 	\\
 Q1054-0020B	& 	0.9514	 	& 	19.28$\pm$0.02 	&	$<-0.21$		& 	$-$1.09$\pm$0.02 	&	$<+0.88$		&	$<+0.52$	&	 	-	 	&	$<+0.36$	&	$<+0.23$		&	 	- 	&	$>$1.68	 	&	$>$1.61		 	&	$>$1.55		\\
 Q1215-0034	& 	1.5543	 	& 	19.56$\pm$0.02 	&	$<-0.56$		& 	$-$0.64$\pm$0.02 	&	$<+0.08$		&	 	-	&	 	-	 	&	$<-0.20$	&	$<-0.79$		&	 	- 	&	-	 	&	-		 	&	-		\\
 Q1220-0040	& 	0.9746	 	& 	20.20$\pm$0.07	&	$<-1.14$		& 	$-$1.33$\pm$0.07 	&	$<+0.19$		&	 	-	&	 	-	 	&	$<-0.14$	&	$<-0.42$        	&	 	- 	&	$>$2.69	 	&	$>$2.19		 	&	 1.72$\pm$0.06	\\
 Q1228+1018	& 	0.9376	 	& 	19.41$\pm$0.02 	&	 $<-0.37$		& 	$-$0.30$\pm$0.02	&	$<-0.07$		&	 	-	&	 	-	 	&	$<-0.55$	&	$<-0.47$	 	&	 	- 	&	$>$2.29	 	&	-		 	&	-		\\
 Q1330-2056	& 	0.8526	 	& 	19.40$\pm$0.02 	&	$<-0.07$		& 	$-$1.07$\pm$0.02 	&	$<+1.00$		&	 	-	&	 	- 		&	$<+0.20$	&	$<-0.22$		&	 	- 	&	$>$1.88	 	&	$>$1.62		 	&	 1.21$\pm$0.05	\\
 Q1436-0051A	& 	0.7377	 	& 	20.08$\pm$0.11 	&	$-$0.05$\pm$0.12 	& 	$-$0.61$\pm$0.11 	& 	+0.58$\pm$0.04	 	&	 	-	&	 $-$0.98$\pm$0.03 	&	$<-0.41$	& 	+0.04$\pm$0.03 		&	 	- 	&	-	 	&	-		 	&	- 		\\
 Q1436-0051B	& 	0.9281	 	& 	$<$18.8		&	$>$+0.86	 	& 	$>-$0.07	 	& 	+0.94$\pm$0.06 		& 	+0.75$\pm$0.05 	&	 $-$0.79$\pm$0.04 	&	$<+0.20$	& 	+0.22$\pm$0.05 		&	 	- 	&	$>$2.63	 	&	$>$1.72		 	&	0.80$\pm$0.03   \\
 Q1455-0045	& 	1.0929	 	& 	20.08$\pm$0.06	&	$<-0.80$		& 	$-$0.98$\pm$0.06 	&	$<+0.18$		& 	+0.00$\pm$0.10 	&	 	-	 	&    $-$0.35$\pm$0.13	& 	-0.51$\pm$0.15 		&	$<-$0.18 	&	$>$2.39	 	&	$>$1.27		 	&	0.92$\pm$0.02	\\
\hline
\end{tabular}
$^{a}$Ratios of column densities
\end{minipage}

\setcounter{table}{14}
\end{sidewaystable*}

\section*{Acknowledgments}
We thank the helpful staff of Las Campanas Observatory for their assistance during the observing runs.
J. Meiring and V.P. Kulkarni gratefully acknowledge support from the National Science 
Foundation grant AST-0607739 (PI Kulkarni). J. Meiring acknowledges partial support from a South Carolina Space Grant graduate student fellowship.

\section*{Appendix - N$_{\rm H \ I}$ determinations}
   The systems studied in this work all have known N$_{\rm H \ I}$ from HST spectra. For completeness and convenience to the reader, we provide plots of the Voigt profiles of the
Lyman-$\alpha$ transition using the best fit values of the column density from \citet{Rao06}. Due to the low resolution and S/N of
these UV spectra, only one component was used in the fits.  We show in figures A1 through A10 the Voigt profiles corresponding 
to the column densities given by \citet{Rao06} and convolved with a Gaussian instrumental spread function based on a two pixel resolution element, superimposed on the
archival data from HST cycle 6 program 6577, cycle 9 program 8569, and cycle 11 program 9382. 
We note that the normalization, i.e., the continuum fit that we define may differ from that adopted by \citet{Rao06}. For our continuum fits, a polynomial 
typically of order 5 or less or a cubic spline was used, and the absorption line itself was excluded from the fitting region.
Also over-plotted are profiles with H I column densities smaller and larger by 0.15 dex than the best fit values. For the \za=0.7377 system in Q1436-0051 in which the \nhI value is 
given as an upper limit, we plot this value (log \nhI=18.8), along with profiles with H I column densities smaller and larger by 0.15 dex.

A horizontal bar in the middle of the profile or directly beneath it denotes the region in velocity space where Mg II absorption was seen. 
This spread in Mg II absorption is likely a lower limit ot the spread in H I, as some components could have detectable H I without detectable Mg II. 
It should be noted that it is possible that \nhI may be overestimated when using only one component to fit the profiles. Multiple components spread over the width of absorption 
seen in Mg II may decrease the total \nhI needed to fit the profile, and thus increase the abundances. 
Higher resolution UV spectra covering higher Lyman series transitions 
would help to resolve the effects of the wide spread of the Mg II components, and discern abundance differences between components which is not
possible with the current lower resolution UV spectra.

\begin{figure*}
\begin{minipage}{160mm}
\begin{center}
$\begin{array}{c@{\hspace{0.5cm}}c}
\multicolumn{1}{l}{\mbox{\bf }} &
	\multicolumn{1}{l}{\mbox{\bf }} \\ 
		\includegraphics[angle=90,width=3.2in,height=2.5in]{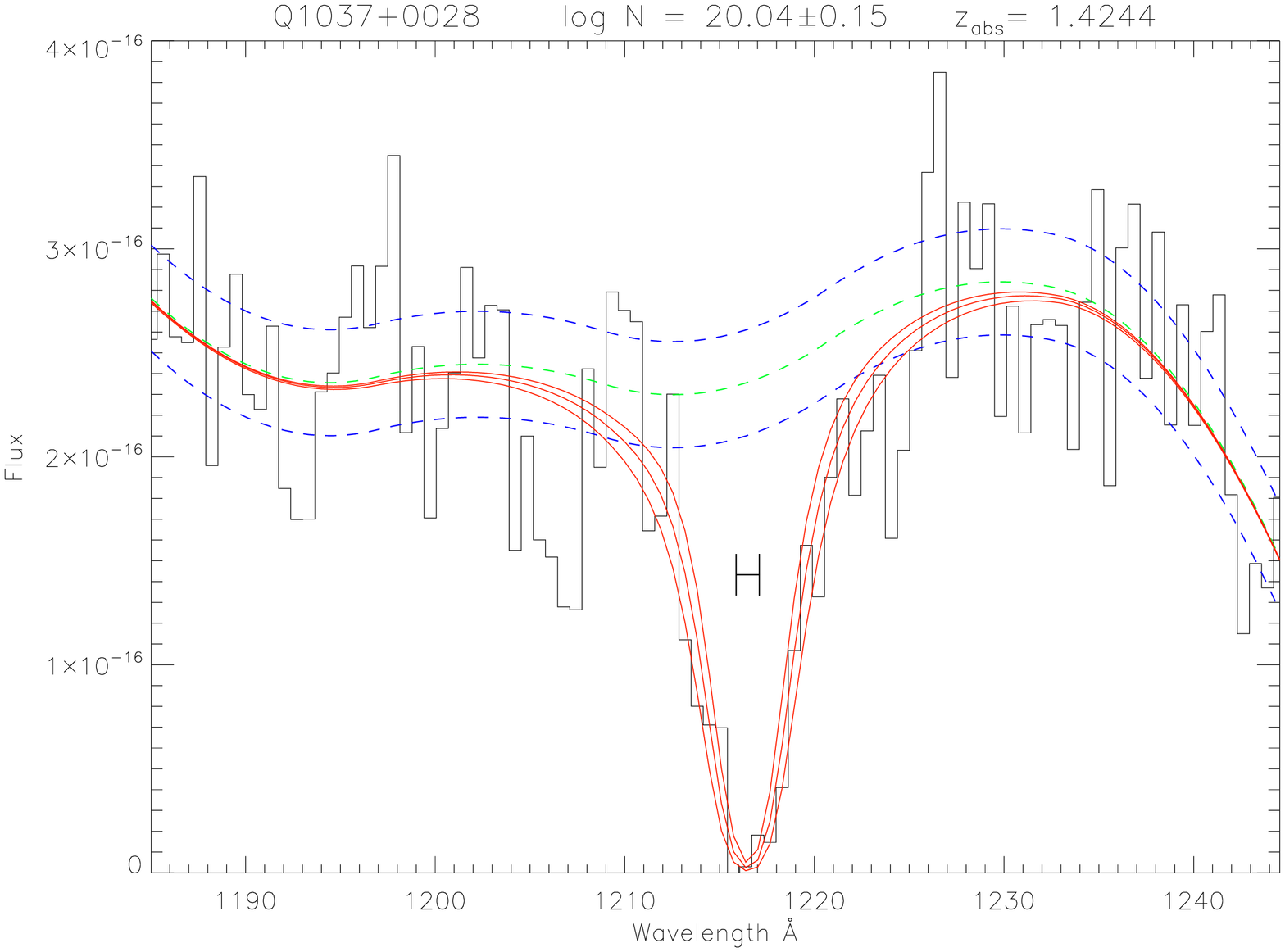} &  \includegraphics[angle=90,width=3.2in,height=2.5in]{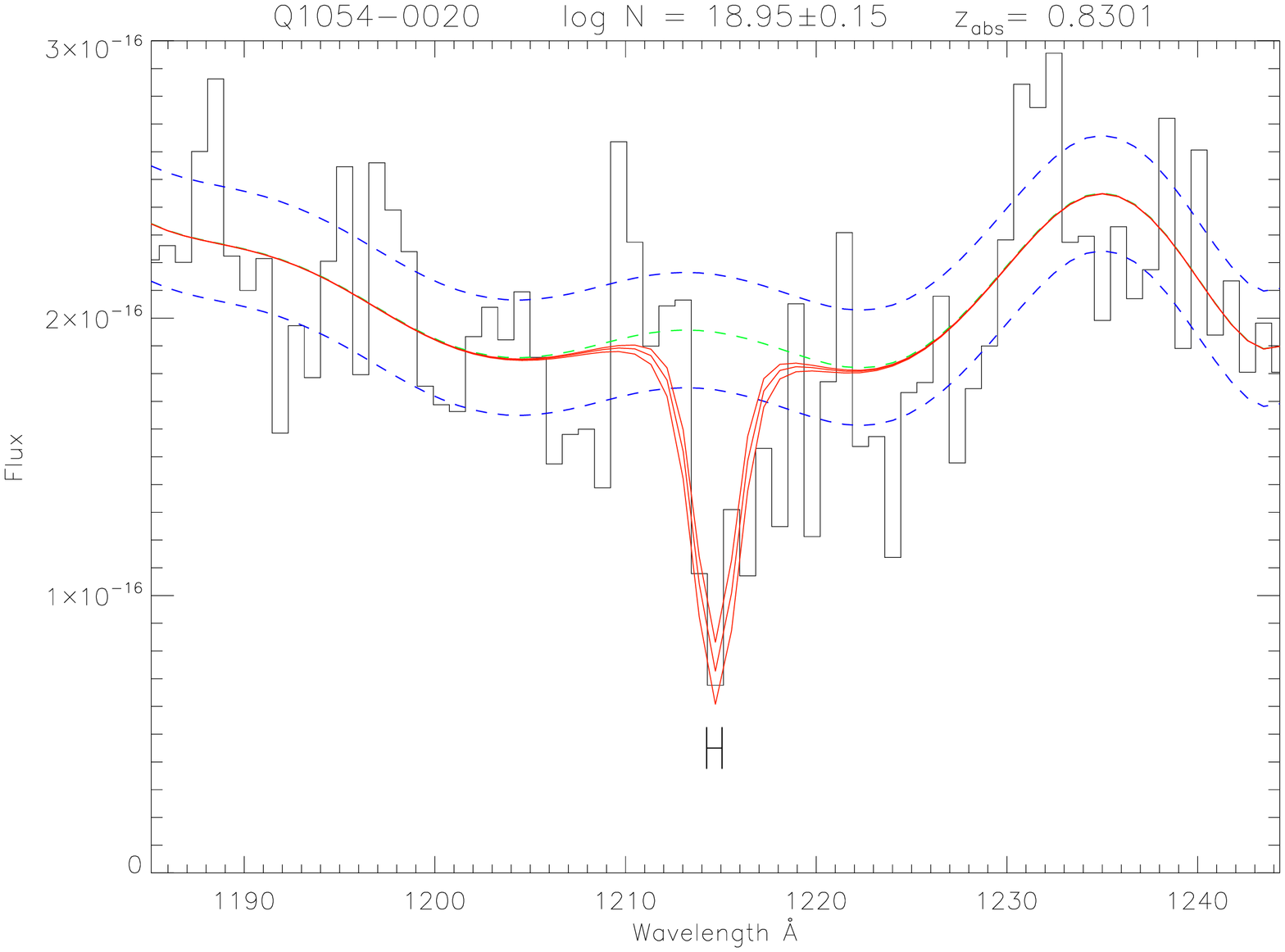} \\ [-0.0cm]
		\textbf{Figure \  A1.} & \textbf{Figure \ A2.} \\
		\includegraphics[angle=90,width=3.2in,height=2.5in]{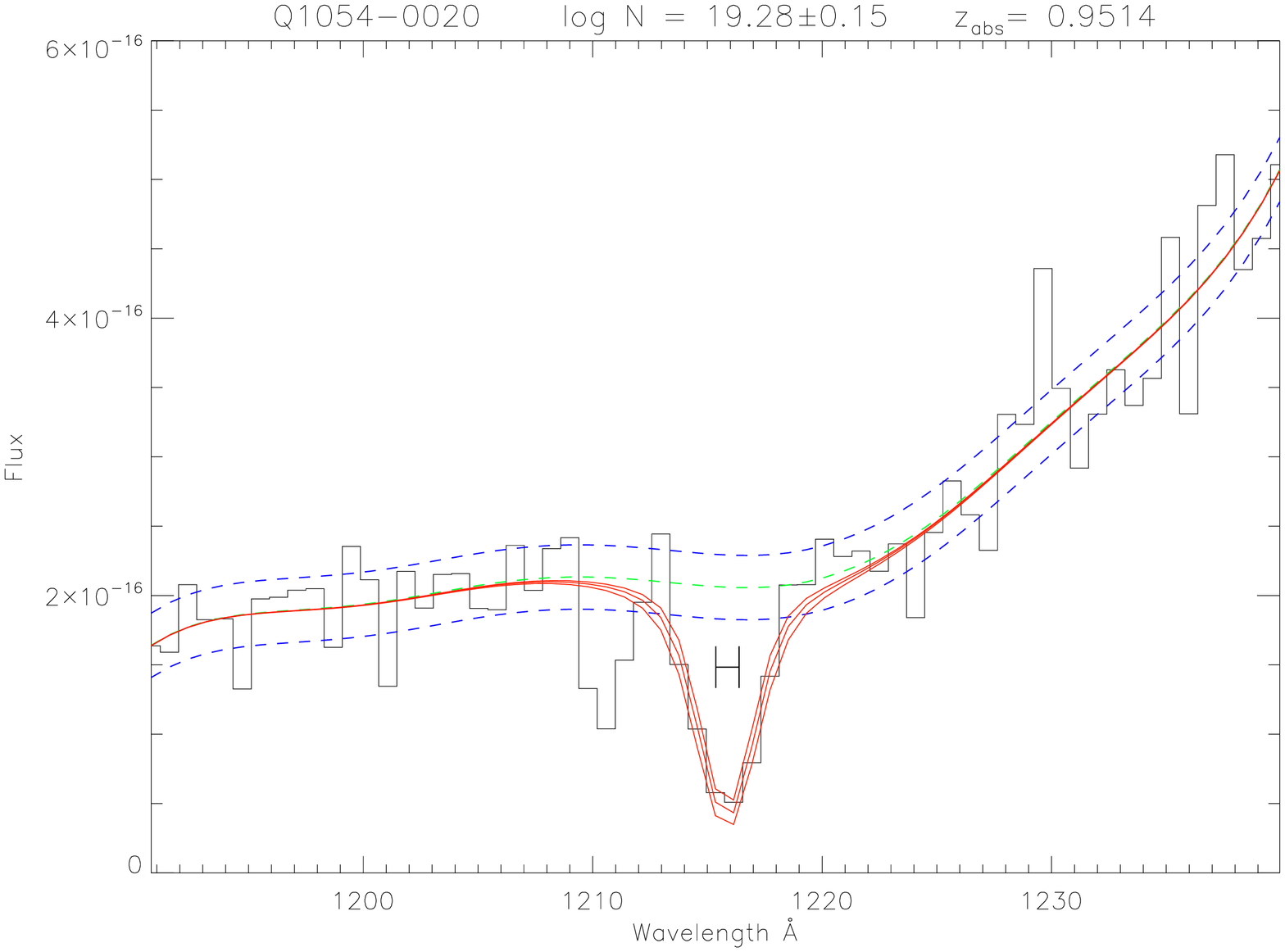} & \includegraphics[angle=90,width=3.2in,height=2.5in]{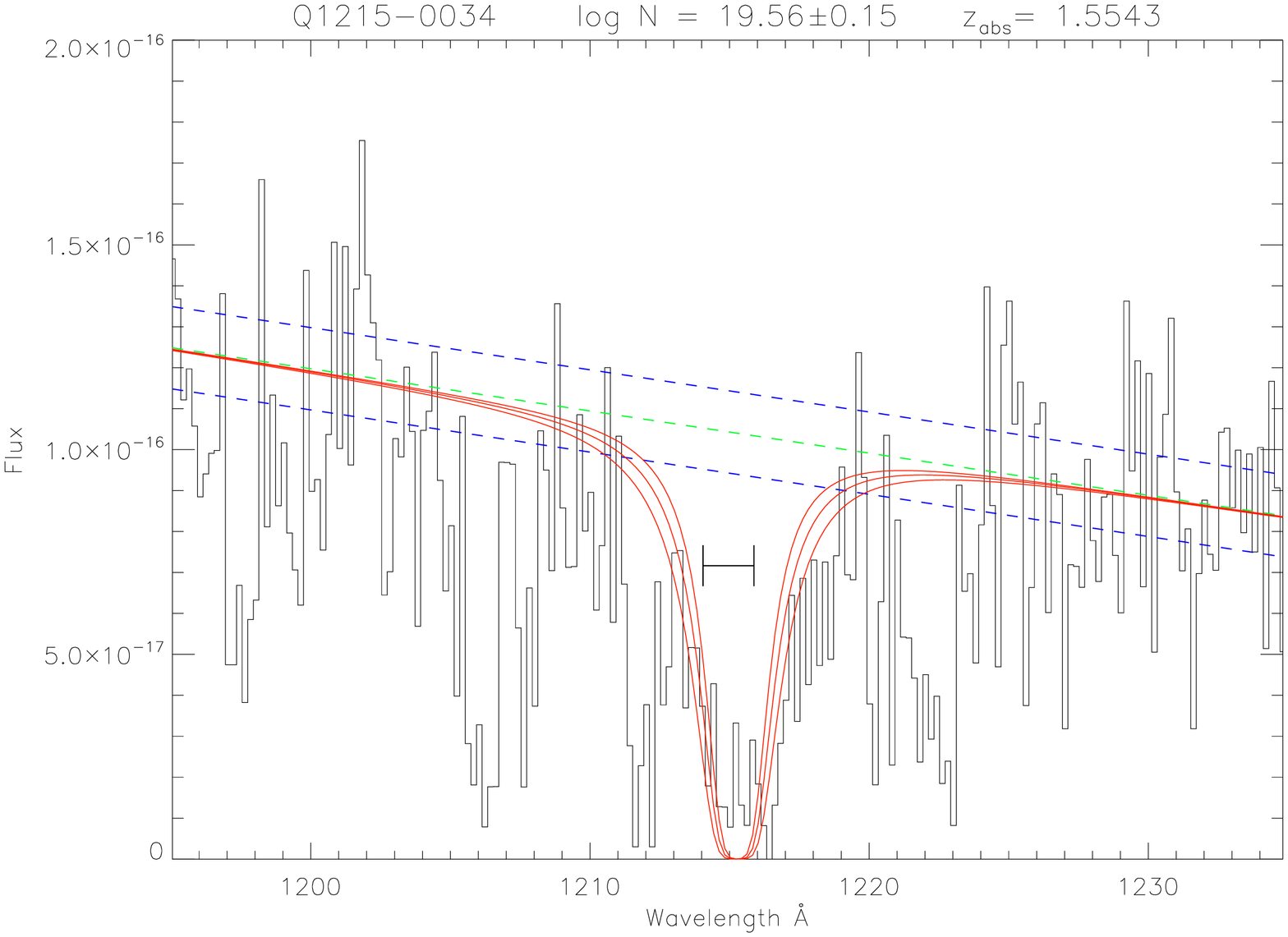} \\  
		\textbf{Figure \ A3.} & \textbf{Figure \ A4.} \\
		\includegraphics[angle=90,width=3.2in,height=2.5in]{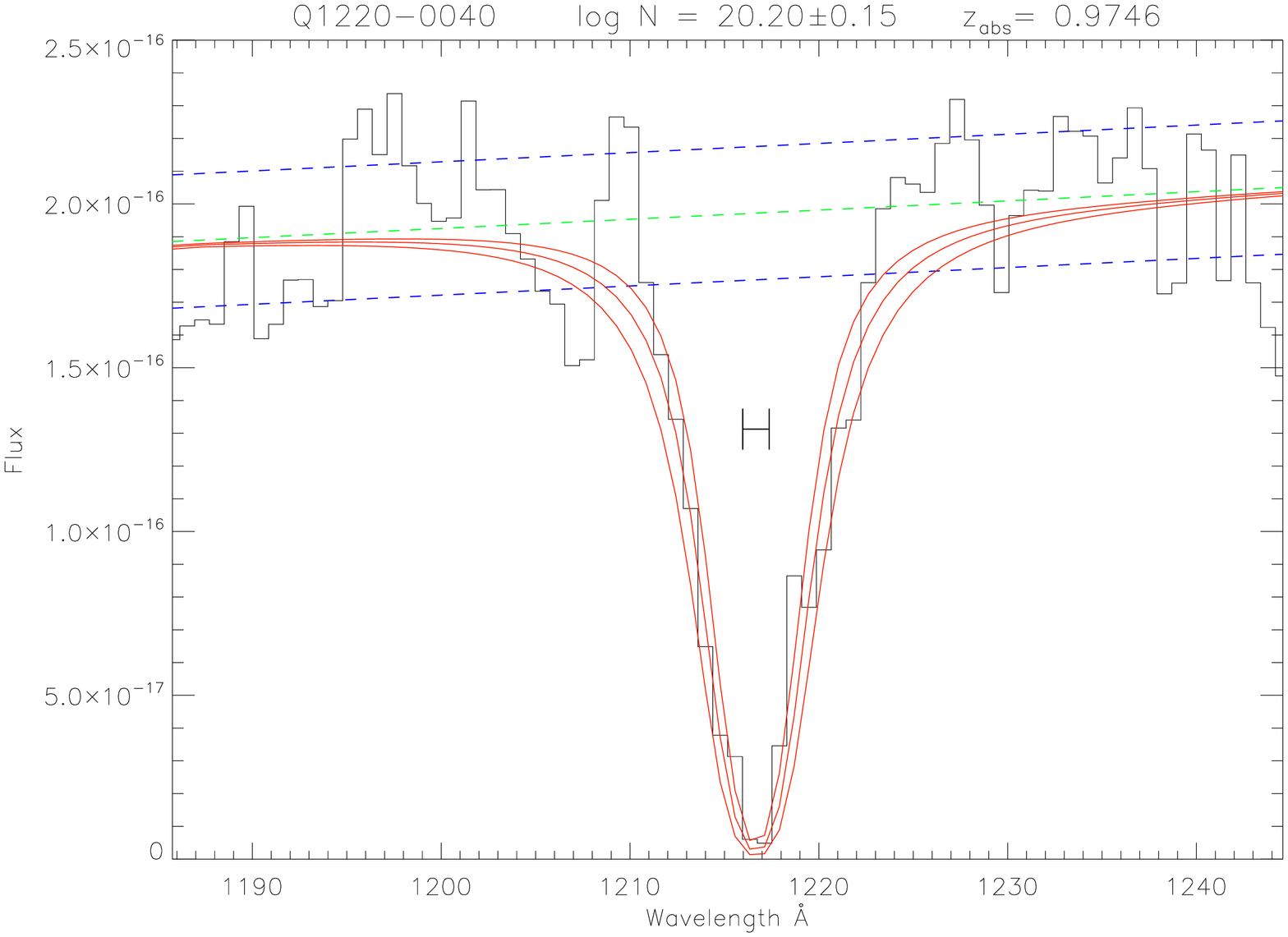} & \includegraphics[angle=90,width=3.2in,height=2.5in]{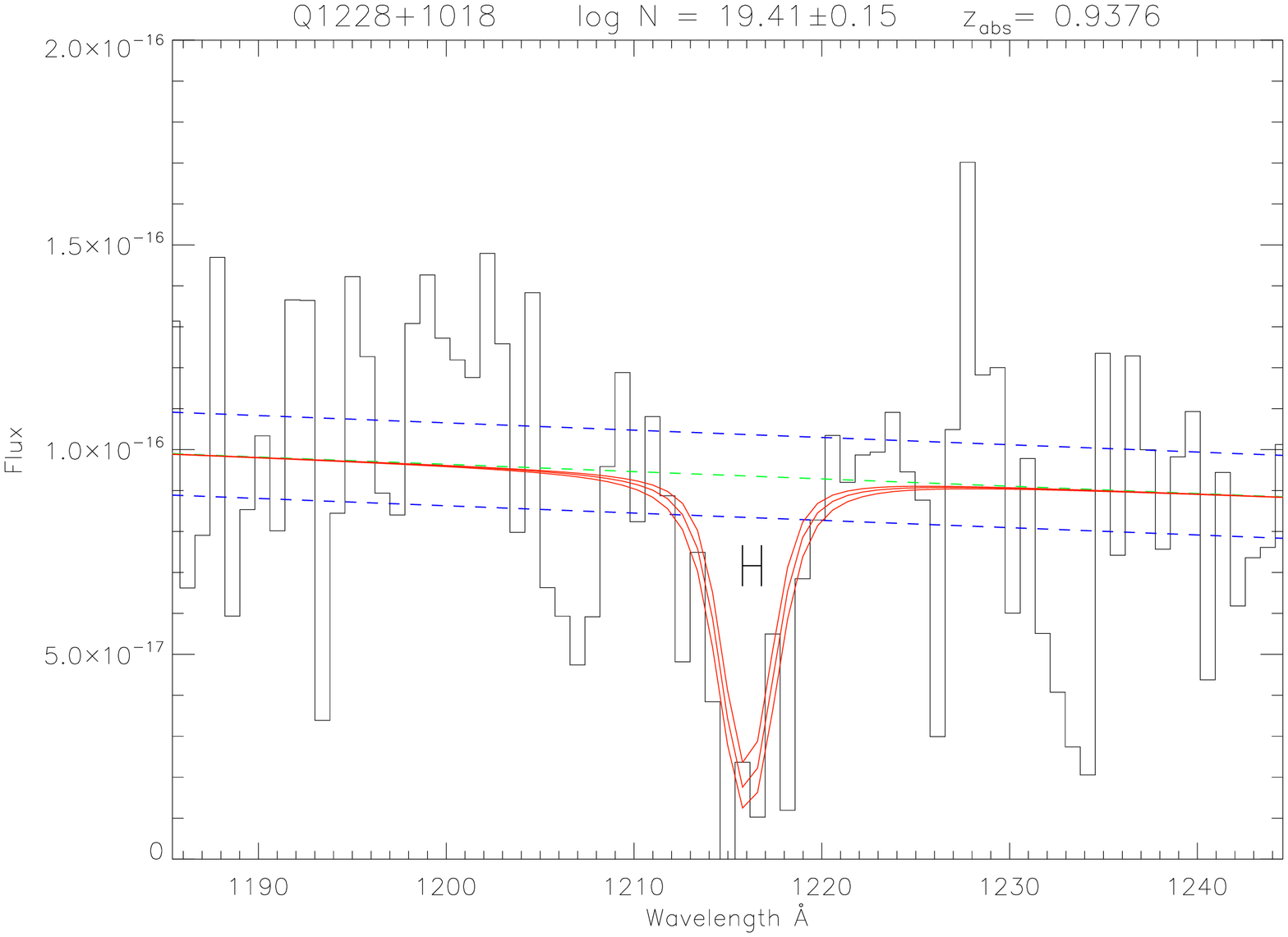} \\  
		\textbf{Figure \  A5.} & \textbf{Figure \ A6.} \\
\end{array}$
\end{center}
\small{\textbf{Figure A1 - A6:}  UV spectra of the systems (clockwise from top left) \za=1.4244 in Q1037+0038, \za=0.8301 in Q1054-0020, \za=1.5543 in Q1215-0034
\za=0.9376 in Q1228+1018, \za=0.9746 in Q1220-0040 and \za=0.9514 in Q1054-0020. 
The dashed green line indicates the best fit level of the continuum, and the dashed blue lines are increased or decreased
by 10$\%$ of the best fit value. Superimposed on the data are theoretical Voigt profiles that have been convolved with a gaussian instrumental spread function. The middle profile is the best fit value from
\citet{Rao06}, and the upper and lower column densities have been modified by $\pm$0.15 dex from the best fit value. The horizontal bar either in the center of the profile or directly 
beneath the line center denotes the region in velocity space where Mg II absorption is seen.} 
\end{minipage}
\end{figure*}

\begin{figure*}
\begin{minipage}{160mm}
\begin{center}
$\begin{array}{c@{\hspace{0.0in}}c}
\multicolumn{1}{l}{\mbox{\bf }} &
	\multicolumn{1}{l}{\mbox{\bf }} \\ [0.0cm]
		\includegraphics[angle=90,width=3.2in,height=2.8in]{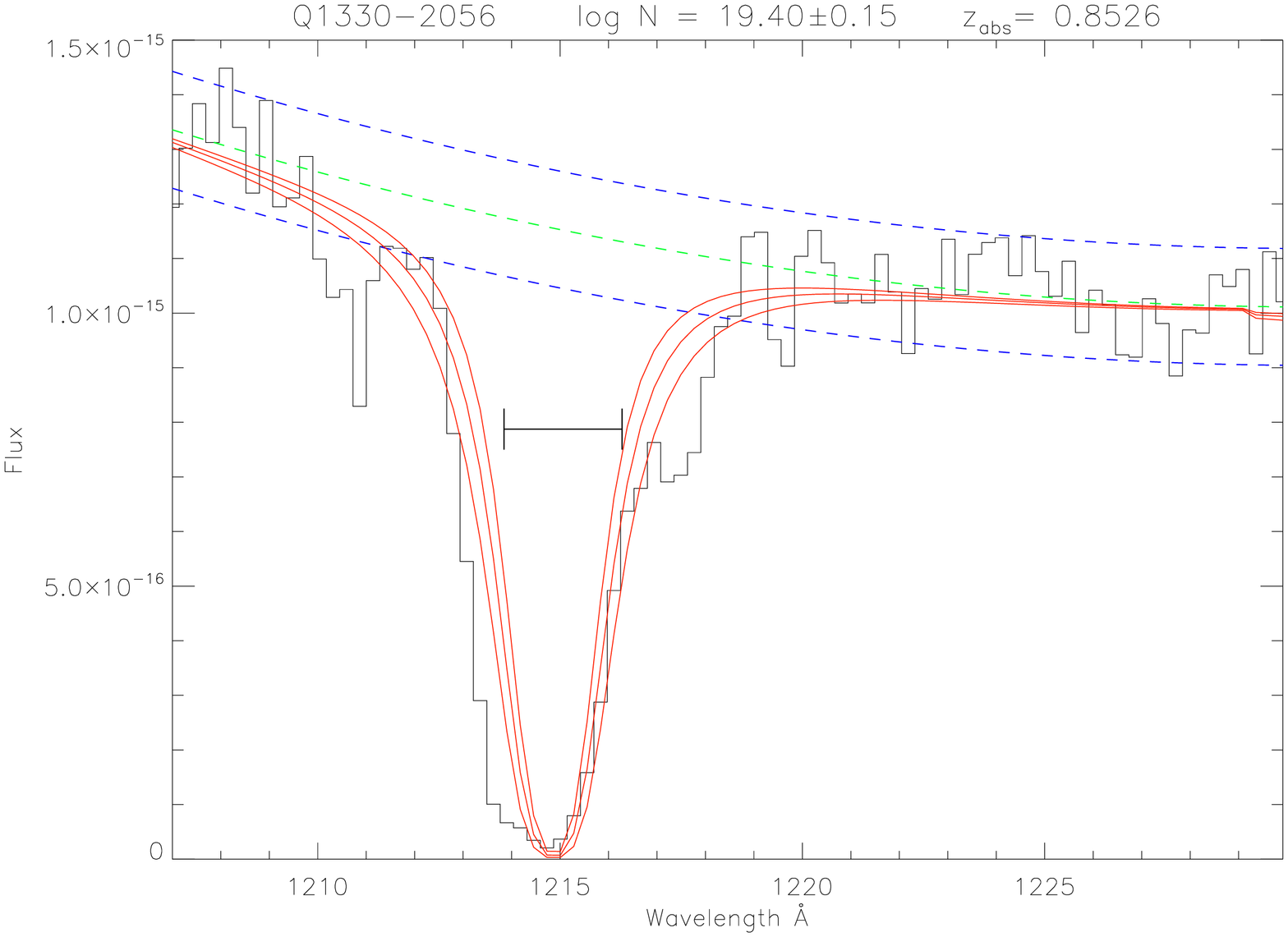} & \includegraphics[angle=90,width=3.2in,height=2.8in]{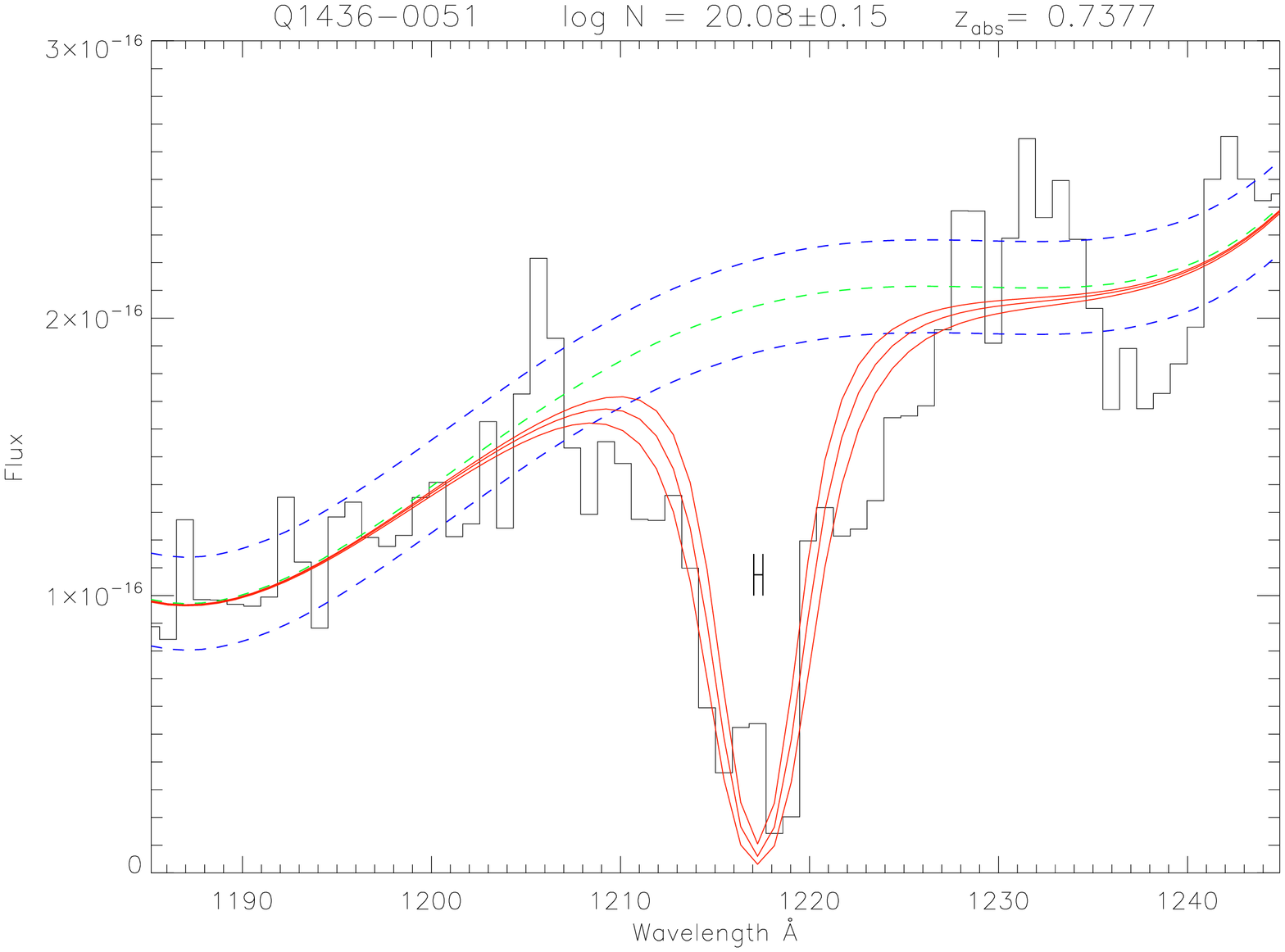} \\ [-0.0cm]
		\textbf{Figure \ A7.} & \textbf{Figure \ A8.} \\ 
		\includegraphics[angle=90,width=3.2in,height=2.8in]{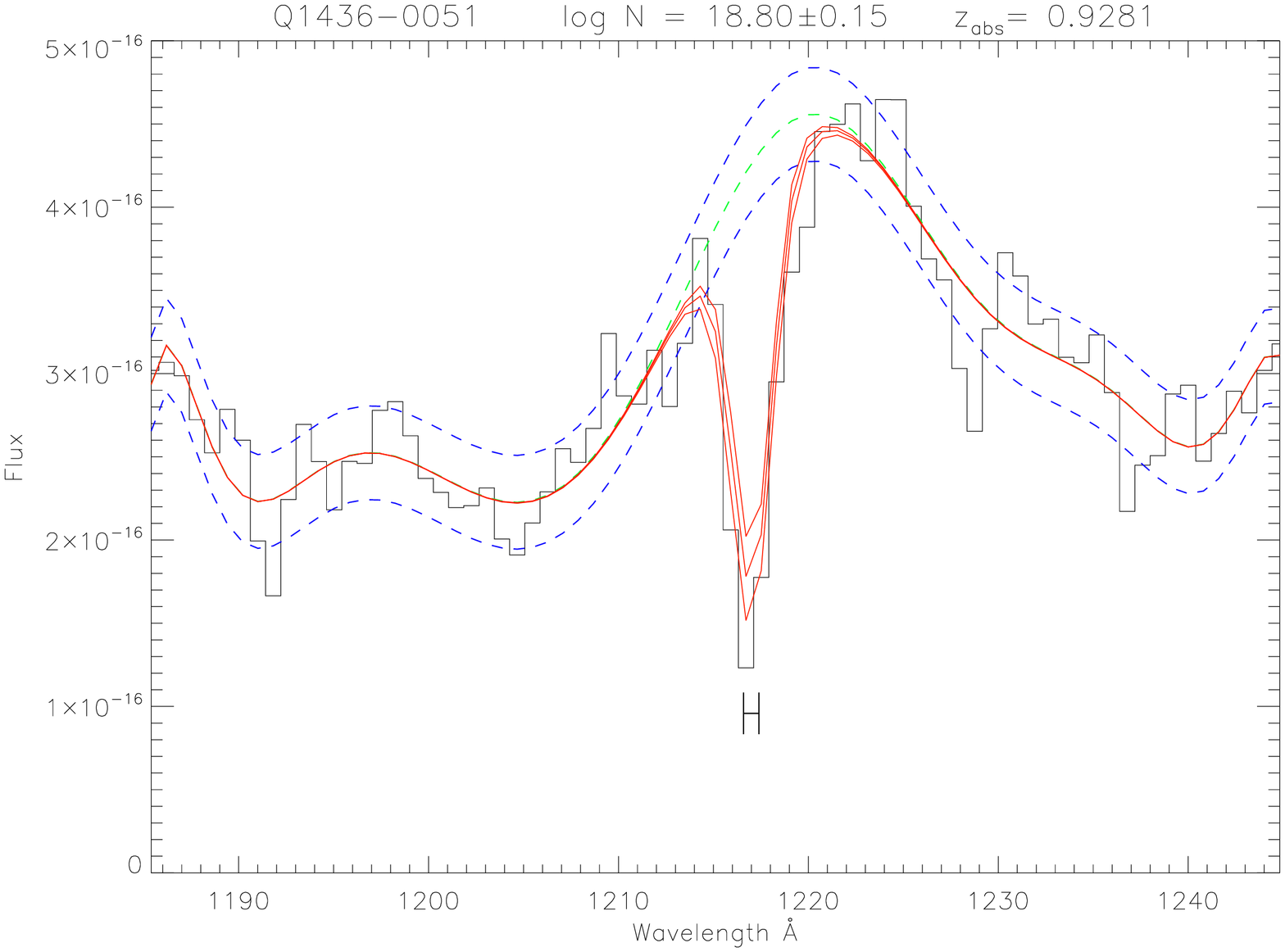} & \includegraphics[angle=90,width=3.2in,height=2.8in]{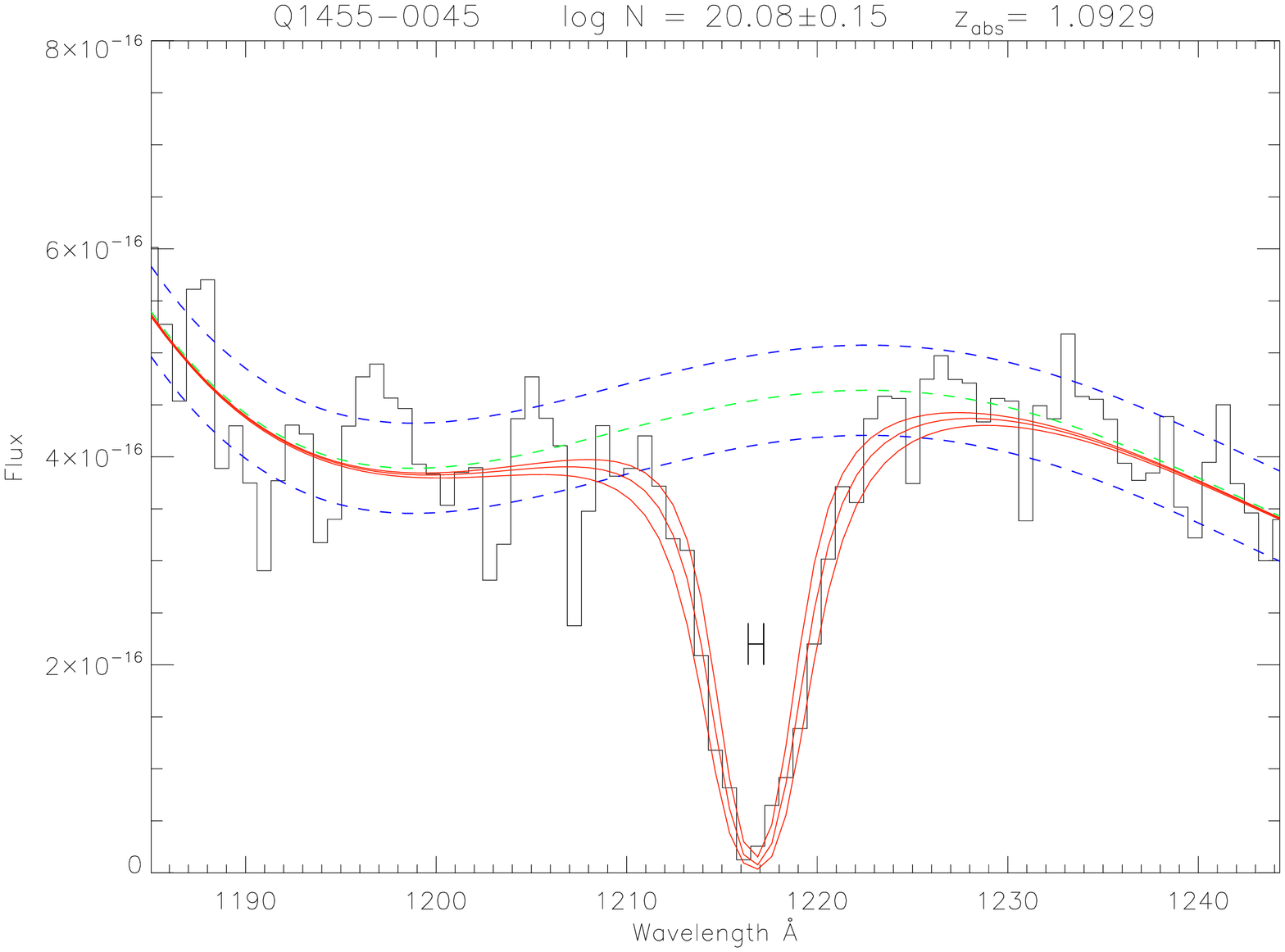} \\  
		\textbf{Figure \ A9.} & \textbf{Figure \ A10.} \\
\end{array}$
\end{center}
\small{\textbf{Figure A7 - A10:} Same as Figures 11-16, but for the systems (clockwise from top left) \za=0.8526 in Q1330-2056, \za=0.7377 in Q1436-0051, \za=1.0929 in 
Q1455-0045, and \za=0.9281 in Q1436-0051.}
\end{minipage}
\end{figure*}

\bsp

\label{lastpage}

\end{document}